\documentclass{article}

\usepackage{jucs2e}
\usepackage{url}
\usepackage[authoryear]{natbib}
\usepackage{bm}
\usepackage{proof}
\usepackage{xcolor}
\usepackage{mathtools}
\usepackage{enumitem}
\makeatletter
\@ifundefined{lhs2tex.lhs2tex.sty.read}%
  {\@namedef{lhs2tex.lhs2tex.sty.read}{}%
   \newcommand\SkipToFmtEnd{}%
   \newcommand\EndFmtInput{}%
   \long\def\SkipToFmtEnd#1\EndFmtInput{}%
  }\SkipToFmtEnd

\newcommand\ReadOnlyOnce[1]{\@ifundefined{#1}{\@namedef{#1}{}}\SkipToFmtEnd}
\usepackage{amstext}
\usepackage{amssymb}
\usepackage{stmaryrd}
\DeclareFontFamily{OT1}{cmtex}{}
\DeclareFontShape{OT1}{cmtex}{m}{n}
  {<5><6><7><8>cmtex8
   <9>cmtex9
   <10><10.95><12><14.4><17.28><20.74><24.88>cmtex10}{}
\DeclareFontShape{OT1}{cmtex}{m}{it}
  {<-> ssub * cmtt/m/it}{}

\DeclareFontShape{OT1}{cmtt}{bx}{n}
  {<5><6><7><8>cmtt8
   <9>cmbtt9
   <10><10.95><12><14.4><17.28><20.74><24.88>cmbtt10}{}
\DeclareFontShape{OT1}{cmtex}{bx}{n}
  {<-> ssub * cmtt/bx/n}{}

\newcommand{\Conid}[1]{\mathit{#1}}
\newcommand{\Varid}[1]{\mathit{#1}}
\newcommand{\anonymous}{\kern0.06em \vbox{\hrule\@width.5em}}
\newcommand{\plus}{\mathbin{+\!\!\!+}}

\renewcommand{\leq}{\leqslant}

\usepackage{polytable}

\@ifundefined{mathindent}%
  {\newdimen\mathindent\mathindent\leftmargini}%
  {}%

\def\resethooks{%
  \global\let\SaveRestoreHook\empty
  \global\let\ColumnHook\empty}
\newcommand*{\savecolumns}[1][default]%
  {\g@addto@macro\SaveRestoreHook{\savecolumns[#1]}}
\newcommand*{\restorecolumns}[1][default]%
  {\g@addto@macro\SaveRestoreHook{\restorecolumns[#1]}}
\newcommand*{\aligncolumn}[2]%
  {\g@addto@macro\ColumnHook{\column{#1}{#2}}}

\resethooks

\newcommand{\onelinecommentchars}{\quad-{}- }
\newcommand{\commentbeginchars}{\enskip\{-}
\newcommand{\commentendchars}{-\}\enskip}

\newcommand{\visiblecomments}{%
  \let\onelinecomment=\onelinecommentchars
  \let\commentbegin=\commentbeginchars
  \let\commentend=\commentendchars}

\newcommand{\invisiblecomments}{%
  \let\onelinecomment=\empty
  \let\commentbegin=\empty
  \let\commentend=\empty}

\visiblecomments

\newlength{\blanklineskip}
\newcommand{\hsindent}[1]{\quad}
\let\hspre\empty
\let\hspost\empty

\EndFmtInput
\makeatother
\ReadOnlyOnce{polycode.fmt}%
\makeatletter

\newcommand{\hsnewpar}[1]%
  {{\parskip=0pt\parindent=0pt\par\vskip #1\noindent}}

\newcommand{\hscodestyle}{}

\newcommand{\sethscode}[1]%
  {\expandafter\let\expandafter\hscode\csname #1\endcsname
   \expandafter\let\expandafter\endhscode\csname end#1\endcsname}

  {\par\noindent
   \advance\leftskip\mathindent
   \hscodestyle
   \let\\=\@normalcr
   \let\hspre\(\let\hspost\)%
   \pboxed}%
  {\endpboxed\)%
   \par\noindent
   \ignorespacesafterend}

  {\hsnewpar\abovedisplayskip
   \advance\leftskip\mathindent
   \hscodestyle
   \let\hspre\(\let\hspost\)%
   \pboxed}%
  {\endpboxed%
   \hsnewpar\belowdisplayskip
   \ignorespacesafterend}

  {\hsnewpar\abovedisplayskip
   \advance\leftskip\mathindent
   \hscodestyle
   \let\\=\@normalcr
   \(\pboxed}%
  {\endpboxed\)%
   \hsnewpar\belowdisplayskip
   \ignorespacesafterend}

\newcommand{\plainhs}{\sethscode{plainhscode}}

\plainhs

  {\hsnewpar\abovedisplayskip
   \advance\leftskip\mathindent
   \hscodestyle
   \let\\=\@normalcr
   \(\parray}%
  {\endparray\)%
   \hsnewpar\belowdisplayskip
   \ignorespacesafterend}

  {\parray}{\endparray}

  {\(\parray}{\endparray\)}

\def\codeframewidth{\arrayrulewidth}
\RequirePackage{calc}

  {\parskip=\abovedisplayskip\par\noindent
   \hscodestyle
   \arrayrulewidth=\codeframewidth
   \tabular{@{}|p{\linewidth-2\arraycolsep-2\arrayrulewidth-2pt}|@{}}%
   \hline\framedhslinecorrect\\{-1.5ex}%
   \let\endoflinesave=\\
   \let\\=\@normalcr
   \(\pboxed}%
  {\endpboxed\)%
   \framedhslinecorrect\endoflinesave{.5ex}\hline
   \endtabular
   \parskip=\belowdisplayskip\par\noindent
   \ignorespacesafterend}

\newcommand{\framedhslinecorrect}[2]%
  {#1[#2]}

  {\(\def\column##1##2{}%
   \let\>\undefined\let\<\undefined\let\\\undefined
   \newcommand\>[1][]{}\newcommand\<[1][]{}\newcommand\\[1][]{}%
   \def\fromto##1##2##3{##3}%
   }{\) }%

  {\let\orighscode=\hscode
   \let\origendhscode=\endhscode
   \def\endhscode{\def\hscode{\endgroup\def\@currenvir{hscode}\\}\begingroup}
   \orighscode\def\hscode{\endgroup\def\@currenvir{hscode}}}%
  {\origendhscode
   \global\let\hscode=\orighscode
   \global\let\endhscode=\origendhscode}%

\makeatother
\EndFmtInput

\newcommand{\inference}[3]{\infer[\mathsf{#2}]{#3}{#1}}
\newcommand{\strutU}{\rule{0pt}{0.5\baselineskip}}
\newcommand{\hlt}[1]{\colorbox{lightgray}{$\displaystyle \strutU{#1}$}}

\newcommand{\nonewlinecode}{\renewcommand{\hsnewpar}[1]%
                                {{\parskip=0pt\parindent=0pt\par\vskip 0pt\noindent}}}

\begin{document}

\title{Trees That Grow\\
       {\color{red}(an early draft -- feedback is sought)}}

\author{{\bfseries Shayan Najd}\\
   (Laboratory for Foundations of Computer Science\\
    The University of Edinburgh, Scotland, U.K.\\
    sh.najd@gmail.com)
   \and
   {\bfseries Simon Peyton Jones}\\
   (Microsoft Research, Cambridge, U.K. \\
   simonpj@microsoft.com)\\
}

\maketitle


\begin{abstract}
We study the notion of extensibility in functional data types, as a new approach
to the problem of decorating abstract syntax trees with additional sets of
information. We observed the need for such extensibility while redesigning the
data types representing Haskell abstract syntax inside GHC.

Specifically, we describe our approach to the tree-decoration problem using a
novel syntactic machinery in Haskell for expressing extensible data types.  We
show that the syntactic machinery is complete in that it can express all the
syntactically possible forms of extensions to algebraic data type
declarations. Then, we describe an encoding of the syntactic machinery based on
the existing features in Glasgow Haskell Compiler (GHC).


\end{abstract}

%

\section{Introduction}
Algebraic Data Types (ADTs) and pattern matching in functional languages lay a
fertile ground to conveniently define and process data types as tree-like
structures. However, in these grounds, trees often cannot grow; once a data type
is defined and compiled, its definition cannot be extended. A data type can be
extended, for instance, by adding new data constructors, and/or by adding new
fields to its existing data constructors.

At the centre of all compilers stand tall trees representing the abstract syntax
of terms.  Compiler programs processing these trees often do so by decorating
the trees with additional information. For instance, name resolution phase adds
information about names, and type inference phase stores the inferred types in
the relevant nodes. We refer to such extra information as decorations.  The
additional information may appear, for instance, as new fields to the existing
data constructors, and/or new data constructors in data types representing the
trees.

Common practice in compilers is either to define a new separate data type
representing the output decorated trees, or to use the same large data type to
represent both the non-decorated input and the decorated output trees. Both
methods are unsatisfactory: the former leads to duplication, and the latter
forces the input trees to carry an unnecessary set of information making them
inconvenient to work with.

We propose a third approach: declare abstract syntax trees with extensible data
types, and view decorations in trees as sets of extensions to the data type
declarations. Extensible data types are the soil in which trees can grow;
extensible data types allow for an arbitrary set of extensions to the same
parametric data type declaration, even after they are compiled.  Our proposed
approach avoids duplication, since the same base declaration is reused for both
non-decorated input trees and the decorated output trees. Since
non-decorated trees are declared by the base (non-extended) data type
declarations, there is no unnecessary set of information baked into the input
trees, making them convenient to work with. Section~\ref{SecDecoratingTrees}
demonstrates the problem and our approach with a running example.

Our proposed approach relies on extensible data types which are less commonly
supported, and often missing as an off-the-shelf feature in functional languages
like Haskell. To be able to adopt this approach in a language like Haskell,
we need to go back to the drawing board, and study the notion of extensibility
for data types in such languages.

Extensions to a data type declaration can appear in different forms.  Earlier,
we enumerated two forms as examples: new fields to the existing data
constructors, and/or new data constructors. We can also consider
extensions to the set of type parameters in a declaration, or in a setting
supporting existential types, we may as well consider extensions to the set of
existentially quantified type variables in the existing data constructors. For
a systematic, yet simple, study, we consider all the syntactically possible
forms of extensions to a generalised algebraic data type (ADT) declaration.  See
Section~\ref{SecAllYouCanDo} for more details about this study.

Having identified different forms of extensions, we describe simple encodings of
extensible data types, within Glasgow Haskell Compiler (GHC), allowing all the
identified forms of extensions by instantiating the same parametric declaration.
In our encodings, extensible data types are parameterised by a set of parameters
representing different forms of extensions, and the act of extending a data type
declaration is simply instantiating these parameters. Setting the same
parameters to the corresponding base cases (e.g., to a type similar to empty
type \ensuremath{\Conid{Void}}, monoidal zero of sum types, or to unit type \ensuremath{()}, monoidal zero of
product types), yields declaration of non-decorated trees. In
Section~\ref{SecGHCCanDo}, we explain the details of such encodings.

\section{Decorating Trees}
\label{SecDecoratingTrees}
In this section, we demonstrates the problem with decorating trees and explain
our solution with a running example.

\subsection{Tree-Decoration Problem}
Consider the following language of lambda terms with integer literals, explicit
type annotations (simple types), tuples (pairs), and let expressions with simple
variable bindings and tuple pattern bindings (projections).

\begin{hscode}\SaveRestoreHook
\column{B}{@{}>{\hspre}l<{\hspost}@{}}%
\column{8}{@{}>{\hspre}l<{\hspost}@{}}%
\column{15}{@{}>{\hspre}l<{\hspost}@{}}%
\column{20}{@{}>{\hspre}c<{\hspost}@{}}%
\column{20E}{@{}l@{}}%
\column{25}{@{}>{\hspre}l<{\hspost}@{}}%
\column{E}{@{}>{\hspre}l<{\hspost}@{}}%
\>[B]{}\Varid{i}\;{}\<[8]%
\>[8]{}\in\;{}\<[15]%
\>[15]{}\text{integers}{}\<[E]%
\\
\>[B]{}\Varid{x},\Varid{y}\;{}\<[8]%
\>[8]{}\in\;{}\<[15]%
\>[15]{}\text{variables}{}\<[E]%
\\
\>[B]{}\Conid{A},\Conid{B},\Conid{C}\;{}\<[8]%
\>[8]{}\in\;{}\<[15]%
\>[15]{}\text{Typ}{}\<[20]%
\>[20]{}\mathbin{::=}{}\<[20E]%
\>[25]{}\textbf{Int}\mid \Conid{A}\boldsymbol{\rightarrow}\Conid{B}\mid \Conid{A}\boldsymbol{\times}\Conid{B}{}\<[E]%
\\
\>[B]{}\Conid{L},\Conid{M},\Conid{N}\;{}\<[8]%
\>[8]{}\in\;{}\<[15]%
\>[15]{}\text{Exp}{}\<[20]%
\>[20]{}\mathbin{::=}{}\<[20E]%
\>[25]{}\Varid{i}\mid \Varid{x}\mid \Conid{M}\!\boldsymbol{::}\!\Conid{A}\mid \boldsymbol{\lambda}\Varid{x}\boldsymbol{.}\,\Conid{N}\mid \Conid{L}\;\Conid{M}\mid \boldsymbol{(}\Conid{M}\boldsymbol{,}\Conid{N}\boldsymbol{)}\mid \textbf{let}\;\Conid{D}\;\textbf{in}\;\Conid{N}{}\<[E]%
\\
\>[B]{}\Conid{D}\;{}\<[8]%
\>[8]{}\in\;{}\<[15]%
\>[15]{}\text{Dec}{}\<[20]%
\>[20]{}\mathbin{::=}{}\<[20E]%
\>[25]{}\Varid{x}\,\boldsymbol{:=}\,\Conid{M}\mid \boldsymbol{(}\Varid{x}\boldsymbol{,}\Varid{y}\boldsymbol{)}\,\boldsymbol{:=}\,\Conid{L}{}\<[E]%
\ColumnHook
\end{hscode}\resethooks

In Haskell, the language above can be declared as the following ADT.

\begin{hscode}\SaveRestoreHook
\column{B}{@{}>{\hspre}l<{\hspost}@{}}%
\column{11}{@{}>{\hspre}c<{\hspost}@{}}%
\column{11E}{@{}l@{}}%
\column{14}{@{}>{\hspre}l<{\hspost}@{}}%
\column{19}{@{}>{\hspre}l<{\hspost}@{}}%
\column{28}{@{}>{\hspre}c<{\hspost}@{}}%
\column{28E}{@{}l@{}}%
\column{31}{@{}>{\hspre}l<{\hspost}@{}}%
\column{36}{@{}>{\hspre}l<{\hspost}@{}}%
\column{41}{@{}>{\hspre}l<{\hspost}@{}}%
\column{46}{@{}>{\hspre}c<{\hspost}@{}}%
\column{46E}{@{}l@{}}%
\column{49}{@{}>{\hspre}l<{\hspost}@{}}%
\column{54}{@{}>{\hspre}l<{\hspost}@{}}%
\column{59}{@{}>{\hspre}l<{\hspost}@{}}%
\column{64}{@{}>{\hspre}c<{\hspost}@{}}%
\column{64E}{@{}l@{}}%
\column{67}{@{}>{\hspre}l<{\hspost}@{}}%
\column{72}{@{}>{\hspre}l<{\hspost}@{}}%
\column{77}{@{}>{\hspre}l<{\hspost}@{}}%
\column{E}{@{}>{\hspre}l<{\hspost}@{}}%
\>[B]{}\mathbf{type}\;\Conid{Var}{}\<[11]%
\>[11]{}\mathrel{=}{}\<[11E]%
\>[14]{}\Conid{String}{}\<[E]%
\\
\>[B]{}\mathbf{data}\;\Conid{Typ}{}\<[11]%
\>[11]{}\mathrel{=}{}\<[11E]%
\>[14]{}\Conid{Int}\mid \Conid{Typ}:\!\rightarrow\Conid{Typ}\mid \Conid{Typ}:\!\!*\!\!:\Conid{Typ}{}\<[E]%
\\
\>[B]{}\mathbf{data}\;\Conid{Exp}{}\<[11]%
\>[11]{}\mathrel{=}{}\<[11E]%
\>[14]{}\Conid{Lit}\;{}\<[19]%
\>[19]{}\Conid{Integer}{}\<[28]%
\>[28]{}\mid {}\<[28E]%
\>[31]{}\Conid{Var}\;{}\<[36]%
\>[36]{}\Conid{Var}{}\<[46]%
\>[46]{}\mid {}\<[46E]%
\>[49]{}\Conid{Typ}\;{}\<[54]%
\>[54]{}\Conid{Exp}\;{}\<[59]%
\>[59]{}\Conid{Typ}{}\<[64]%
\>[64]{}\mid {}\<[64E]%
\>[67]{}\Conid{Abs}\;{}\<[72]%
\>[72]{}\Conid{Var}\;{}\<[77]%
\>[77]{}\Conid{Exp}{}\<[E]%
\\
\>[28]{}\mid {}\<[28E]%
\>[31]{}\Conid{App}\;{}\<[36]%
\>[36]{}\Conid{Exp}\;{}\<[41]%
\>[41]{}\Conid{Exp}{}\<[46]%
\>[46]{}\mid {}\<[46E]%
\>[49]{}\Conid{Tup}\;{}\<[54]%
\>[54]{}\Conid{Exp}\;{}\<[59]%
\>[59]{}\Conid{Exp}{}\<[64]%
\>[64]{}\mid {}\<[64E]%
\>[67]{}\Conid{Let}\;{}\<[72]%
\>[72]{}\Conid{Dec}\;{}\<[77]%
\>[77]{}\Conid{Exp}{}\<[E]%
\\
\>[B]{}\mathbf{data}\;\Conid{Dec}{}\<[11]%
\>[11]{}\mathrel{=}{}\<[11E]%
\>[14]{}\Conid{Val}\;{}\<[19]%
\>[19]{}\Conid{Var}\;\Conid{Exp}{}\<[28]%
\>[28]{}\mid {}\<[28E]%
\>[31]{}\Conid{Prj}\;{}\<[36]%
\>[36]{}\Conid{Var}\;{}\<[41]%
\>[41]{}\Conid{Var}\;\Conid{Exp}{}\<[E]%
\ColumnHook
\end{hscode}\resethooks

Defining a simple printer for this data type is straightforward:

\begin{hscode}\SaveRestoreHook
\column{B}{@{}>{\hspre}l<{\hspost}@{}}%
\column{9}{@{}>{\hspre}l<{\hspost}@{}}%
\column{13}{@{}>{\hspre}c<{\hspost}@{}}%
\column{13E}{@{}l@{}}%
\column{16}{@{}>{\hspre}l<{\hspost}@{}}%
\column{18}{@{}>{\hspre}l<{\hspost}@{}}%
\column{19}{@{}>{\hspre}l<{\hspost}@{}}%
\column{22}{@{}>{\hspre}c<{\hspost}@{}}%
\column{22E}{@{}l@{}}%
\column{23}{@{}>{\hspre}c<{\hspost}@{}}%
\column{23E}{@{}l@{}}%
\column{25}{@{}>{\hspre}l<{\hspost}@{}}%
\column{26}{@{}>{\hspre}l<{\hspost}@{}}%
\column{32}{@{}>{\hspre}l<{\hspost}@{}}%
\column{34}{@{}>{\hspre}l<{\hspost}@{}}%
\column{45}{@{}>{\hspre}c<{\hspost}@{}}%
\column{45E}{@{}l@{}}%
\column{47}{@{}>{\hspre}c<{\hspost}@{}}%
\column{47E}{@{}l@{}}%
\column{49}{@{}>{\hspre}l<{\hspost}@{}}%
\column{51}{@{}>{\hspre}l<{\hspost}@{}}%
\column{57}{@{}>{\hspre}c<{\hspost}@{}}%
\column{57E}{@{}l@{}}%
\column{61}{@{}>{\hspre}c<{\hspost}@{}}%
\column{61E}{@{}l@{}}%
\column{62}{@{}>{\hspre}l<{\hspost}@{}}%
\column{65}{@{}>{\hspre}l<{\hspost}@{}}%
\column{75}{@{}>{\hspre}l<{\hspost}@{}}%
\column{E}{@{}>{\hspre}l<{\hspost}@{}}%
\>[B]{}\Varid{printT}\mathbin{::}\Conid{Typ}\to \Conid{String}{}\<[E]%
\\
\>[B]{}\Varid{printT}\;{}\<[9]%
\>[9]{}\Conid{Int}{}\<[23]%
\>[23]{}\mathrel{=}{}\<[23E]%
\>[26]{}\text{\tt \char34 Int\char34}{}\<[E]%
\\
\>[B]{}\Varid{printT}\;{}\<[9]%
\>[9]{}(\Varid{a}{}\<[13]%
\>[13]{}:\!\rightarrow{}\<[13E]%
\>[18]{}\Varid{b}){}\<[23]%
\>[23]{}\mathrel{=}{}\<[23E]%
\>[26]{}\text{\tt \char34 (\char34}{}\<[32]%
\>[32]{}\plus \Varid{printT}\;\Varid{a}{}\<[45]%
\>[45]{}\plus {}\<[45E]%
\>[49]{}\text{\tt \char34 )~→~\char34}{}\<[57]%
\>[57]{}\plus {}\<[57E]%
\>[62]{}\Varid{printT}\;\Varid{b}{}\<[E]%
\\
\>[B]{}\Varid{printT}\;{}\<[9]%
\>[9]{}(\Varid{a}{}\<[13]%
\>[13]{}:\!\!*\!\!:{}\<[13E]%
\>[18]{}\Varid{b}){}\<[23]%
\>[23]{}\mathrel{=}{}\<[23E]%
\>[26]{}\text{\tt \char34 (\char34}{}\<[32]%
\>[32]{}\plus \Varid{printT}\;\Varid{a}{}\<[45]%
\>[45]{}\plus {}\<[45E]%
\>[49]{}\text{\tt \char34 )~×~\char34}{}\<[57]%
\>[57]{}\plus {}\<[57E]%
\>[62]{}\Varid{printT}\;\Varid{b}{}\<[E]%
\\[\blanklineskip]%
\>[B]{}\Varid{printE}\mathbin{::}\Conid{Exp}\to \Conid{String}{}\<[E]%
\\
\>[B]{}\Varid{printE}\;{}\<[9]%
\>[9]{}(\Conid{Lit}\;{}\<[16]%
\>[16]{}\Varid{i}){}\<[23]%
\>[23]{}\mathrel{=}{}\<[23E]%
\>[26]{}\Varid{show}\;\Varid{i}{}\<[E]%
\\
\>[B]{}\Varid{printE}\;{}\<[9]%
\>[9]{}(\Conid{Var}\;{}\<[16]%
\>[16]{}\Varid{x}){}\<[23]%
\>[23]{}\mathrel{=}{}\<[23E]%
\>[26]{}\Varid{x}{}\<[E]%
\\
\>[B]{}\Varid{printE}\;{}\<[9]%
\>[9]{}(\Conid{Typ}\;{}\<[16]%
\>[16]{}\Varid{m}\;{}\<[19]%
\>[19]{}\Varid{a}){}\<[23]%
\>[23]{}\mathrel{=}{}\<[23E]%
\>[26]{}\text{\tt \char34 (\char34}{}\<[34]%
\>[34]{}\plus \Varid{printE}\;\Varid{m}{}\<[47]%
\>[47]{}\plus {}\<[47E]%
\>[51]{}\text{\tt \char34 )~::~(\char34}{}\<[61]%
\>[61]{}\plus {}\<[61E]%
\>[65]{}\Varid{printT}\;\Varid{a}{}\<[75]%
\>[75]{}\plus \text{\tt \char34 )\char34}{}\<[E]%
\\
\>[B]{}\Varid{printE}\;{}\<[9]%
\>[9]{}(\Conid{Abs}\;{}\<[16]%
\>[16]{}\Varid{x}\;{}\<[19]%
\>[19]{}\Varid{n}){}\<[23]%
\>[23]{}\mathrel{=}{}\<[23E]%
\>[26]{}\text{\tt \char34}\lambda\text{\tt \char34}{}\<[34]%
\>[34]{}\plus \Varid{x}\plus \text{\tt \char34 .\char34}\plus \Varid{printE}\;\Varid{n}{}\<[E]%
\\
\>[B]{}\Varid{printE}\;{}\<[9]%
\>[9]{}(\Conid{App}\;{}\<[16]%
\>[16]{}\Varid{l}\;{}\<[19]%
\>[19]{}\Varid{m}){}\<[23]%
\>[23]{}\mathrel{=}{}\<[23E]%
\>[26]{}\text{\tt \char34 (\char34}{}\<[34]%
\>[34]{}\plus \Varid{printE}\;\Varid{l}{}\<[47]%
\>[47]{}\plus {}\<[47E]%
\>[51]{}\text{\tt \char34 )~(\char34}{}\<[61]%
\>[61]{}\plus {}\<[61E]%
\>[65]{}\Varid{printE}\;\Varid{m}{}\<[75]%
\>[75]{}\plus \text{\tt \char34 )\char34}{}\<[E]%
\\
\>[B]{}\Varid{printE}\;{}\<[9]%
\>[9]{}(\Conid{Tup}\;{}\<[16]%
\>[16]{}\Varid{m}\;{}\<[19]%
\>[19]{}\Varid{n}){}\<[23]%
\>[23]{}\mathrel{=}{}\<[23E]%
\>[26]{}\text{\tt \char34 (\char34}{}\<[34]%
\>[34]{}\plus \Varid{printE}\;\Varid{m}{}\<[47]%
\>[47]{}\plus {}\<[47E]%
\>[51]{}\text{\tt \char34 ~,~\char34}{}\<[61]%
\>[61]{}\plus {}\<[61E]%
\>[65]{}\Varid{printE}\;\Varid{n}{}\<[75]%
\>[75]{}\plus \text{\tt \char34 )\char34}{}\<[E]%
\\
\>[B]{}\Varid{printE}\;{}\<[9]%
\>[9]{}(\Conid{Let}\;{}\<[16]%
\>[16]{}\Varid{d}\;{}\<[19]%
\>[19]{}\Varid{n}){}\<[23]%
\>[23]{}\mathrel{=}{}\<[23E]%
\>[26]{}\text{\tt \char34 let~\char34}{}\<[34]%
\>[34]{}\plus \Varid{printD}\;\Varid{d}{}\<[47]%
\>[47]{}\plus {}\<[47E]%
\>[51]{}\text{\tt \char34 ~in~\char34}{}\<[61]%
\>[61]{}\plus {}\<[61E]%
\>[65]{}\Varid{printE}\;\Varid{n}{}\<[E]%
\\[\blanklineskip]%
\>[B]{}\Varid{printD}\mathbin{::}\Conid{Dec}\to \Conid{String}{}\<[E]%
\\
\>[B]{}\Varid{printD}\;{}\<[9]%
\>[9]{}(\Conid{Val}\;\Varid{x}\;\Varid{m}){}\<[22]%
\>[22]{}\mathrel{=}{}\<[22E]%
\>[25]{}\Varid{x}\plus \text{\tt \char34 ~:=~\char34}\plus \Varid{printE}\;\Varid{m}{}\<[E]%
\\
\>[B]{}\Varid{printD}\;{}\<[9]%
\>[9]{}(\Conid{Prj}\;\Varid{x}\;\Varid{y}\;\Varid{l}){}\<[22]%
\>[22]{}\mathrel{=}{}\<[22E]%
\>[25]{}\text{\tt \char34 (\char34}\plus \Varid{x}\plus \text{\tt \char34 ~,~\char34}\plus \Varid{y}\plus \text{\tt \char34 )~:=~\char34}\plus \Varid{printE}\;\Varid{l}{}\<[E]%
\ColumnHook
\end{hscode}\resethooks

Now consider the following standard type system for the above language, with
$\Gamma$ and $\Delta$ ranging over standard type environments.
\[
\begin{array}{@{}l@{}}
\begin{array}{@{}l@{}}
\fbox{$\Gamma \vdash M:A$}\\~\\
\end{array}
\
\inference{}
          {}
          {\Gamma \vdash \ensuremath{\Varid{i}}:\ensuremath{\textbf{Int}}}
\
\inference{(\ensuremath{\Varid{x}}:A) \in \Gamma}
          {}
          {\Gamma \vdash \ensuremath{\Varid{x}}:\ensuremath{\Conid{A}}}
\
\inference{\Gamma \vdash \ensuremath{\Conid{M}}:\ensuremath{\Conid{A}}}
          {}
          {\Gamma \vdash \ensuremath{\Conid{M}\!\boldsymbol{::}\!\Conid{A}}:\ensuremath{\Conid{A}}}
\
\inference{\Gamma \vdash \ensuremath{\Conid{M}} : \ensuremath{\Conid{A}} &
           \Gamma \vdash \ensuremath{\Conid{N}} : \ensuremath{\Conid{B}}}
          {}
          {\Gamma \vdash \ensuremath{\boldsymbol{(}\Conid{M}\boldsymbol{,}\Conid{N}\boldsymbol{)}} : \ensuremath{\Conid{A}\boldsymbol{\times}\Conid{B}}}
\\
\inference{x:A, \Gamma \vdash \ensuremath{\Conid{N}} : \ensuremath{\Conid{B}} }
          {}
          {\Gamma \vdash \ensuremath{\boldsymbol{\lambda}\Varid{x}\boldsymbol{.}\,\Conid{N}} : \ensuremath{\Conid{A}\boldsymbol{\rightarrow}\Conid{B}}}
\
\inference{\Gamma \vdash \ensuremath{\Conid{L}} : \ensuremath{\Conid{A}\boldsymbol{\rightarrow}\Conid{B}} &
           \Gamma \vdash \ensuremath{\Conid{M}} : \ensuremath{\Conid{A}}}
          {}
          {\Gamma \vdash \ensuremath{\Conid{L}\;\Conid{M}} : \ensuremath{\Conid{B}}}
\
\inference{\Gamma \vdash \ensuremath{\Conid{D}} \leadsto \Delta & \Delta, \Gamma \vdash \ensuremath{\Conid{N}} :\ensuremath{\Conid{A}}}
          {}
          {\Gamma \vdash \ensuremath{\textbf{let}\;\Conid{D}\;\textbf{in}\;\Conid{N}} : \ensuremath{\Conid{A}}}
\\
\begin{array}{@{}l@{}}
\fbox{$\Gamma \vdash D \leadsto \Delta$}\\~\\
\end{array}
\
\inference{\Gamma \vdash \ensuremath{\Conid{M}} : \ensuremath{\Conid{A}}}
          {}
          {\Gamma \vdash \ensuremath{\Varid{x}\,\boldsymbol{:=}\,\Conid{M}} \leadsto [ \ensuremath{\Varid{x}}: \ensuremath{\Conid{A}} ]}
\
\inference{\Gamma \vdash \ensuremath{\Conid{L}} : \ensuremath{\Conid{A}\boldsymbol{\times}\Conid{B}}}
          {}
          {\Gamma \vdash \ensuremath{\boldsymbol{(}\Varid{x}\boldsymbol{,}\Varid{y}\boldsymbol{)}\,\boldsymbol{:=}\,\Conid{L}} \leadsto [ \ensuremath{\Varid{x}} : \ensuremath{\Conid{A}}, \ensuremath{\Varid{y}} : \ensuremath{\Conid{B}} ]}
\end{array}
\]

Before type checking, often abstract syntax trees (ASTs) are processed by a type
inference engine. The output of the type inference engine is the same input tree
decorated with additional type information. Type inference helps users to leave
certain bits of their programs without explicit type annotations. Type inference
also helps in simplifying the type checker: after type inference, and decorating
the trees with the additional type information, type checking becomes a
straightforward syntax-directed recursive definition. To accommodate for the
additional information in the output, we need larger trees, and hence we need to
\emph{extend} the original declarations.
For instance, the following highlights the required changes to the \ensuremath{\Conid{Exp}} data
type (besides the trivial updates of \ensuremath{\Conid{Dec}} to \ensuremath{\Conid{Dec}^\bullet} and \ensuremath{\Conid{Exp}} to \ensuremath{\Conid{Exp}^\bullet}).

\begin{hscode}\SaveRestoreHook
\column{B}{@{}>{\hspre}l<{\hspost}@{}}%
\column{12}{@{}>{\hspre}c<{\hspost}@{}}%
\column{12E}{@{}l@{}}%
\column{15}{@{}>{\hspre}l<{\hspost}@{}}%
\column{21}{@{}>{\hspre}l<{\hspost}@{}}%
\column{27}{@{}>{\hspre}l<{\hspost}@{}}%
\column{33}{@{}>{\hspre}c<{\hspost}@{}}%
\column{33E}{@{}l@{}}%
\column{36}{@{}>{\hspre}l<{\hspost}@{}}%
\column{42}{@{}>{\hspre}l<{\hspost}@{}}%
\column{47}{@{}>{\hspre}l<{\hspost}@{}}%
\column{72}{@{}>{\hspre}l<{\hspost}@{}}%
\column{78}{@{}>{\hspre}l<{\hspost}@{}}%
\column{E}{@{}>{\hspre}l<{\hspost}@{}}%
\>[B]{}\mathbf{type}\;\Conid{TypEnv}\mathrel{=}[\mskip1.5mu (\Conid{Var},\Conid{Typ})\mskip1.5mu]{}\<[E]%
\\
\>[B]{}\mathbf{data}\;\Conid{Exp}^\bullet{}\<[12]%
\>[12]{}\mathrel{=}{}\<[12E]%
\>[15]{}\Conid{Lit}^\bullet\;{}\<[21]%
\>[21]{}\Conid{Integer}{}\<[33]%
\>[33]{}\mid {}\<[33E]%
\>[36]{}\Conid{Var}^\bullet\;{}\<[42]%
\>[42]{}\Conid{Var}\mid \Conid{Typ}^\bullet\;{}\<[72]%
\>[72]{}\Conid{Exp}^\bullet\;{}\<[78]%
\>[78]{}\Conid{Typ}{}\<[E]%
\\
\>[12]{}\mid {}\<[12E]%
\>[15]{}\Conid{Abs}^\bullet\;{}\<[21]%
\>[21]{}\Conid{Var}\;{}\<[27]%
\>[27]{}\Conid{Exp}^\bullet{}\<[33]%
\>[33]{}\mid {}\<[33E]%
\>[36]{}\Conid{App}^\bullet\;{}\<[42]%
\>[42]{}\hlt{\Conid{Typ}\;\!}\;{}\<[72]%
\>[72]{}\Conid{Exp}^\bullet\;{}\<[78]%
\>[78]{}\Conid{Exp}^\bullet{}\<[E]%
\\
\>[12]{}\mid {}\<[12E]%
\>[15]{}\Conid{Tup}^\bullet\;{}\<[21]%
\>[21]{}\Conid{Exp}^\bullet\;{}\<[27]%
\>[27]{}\Conid{Exp}^\bullet{}\<[33]%
\>[33]{}\mid {}\<[33E]%
\>[36]{}\Conid{Let}^\bullet\;{}\<[42]%
\>[42]{}\hlt{\Conid{TypEnv}\;\!}\;{}\<[72]%
\>[72]{}\Conid{Dec}^\bullet\;{}\<[78]%
\>[78]{}\Conid{Exp}^\bullet{}\<[E]%
\\
\>[B]{}\mathbf{data}\;\Conid{Dec}^\bullet{}\<[12]%
\>[12]{}\mathrel{=}{}\<[12E]%
\>[15]{}\Conid{Val}^\bullet\;{}\<[21]%
\>[21]{}\Conid{Var}\;{}\<[27]%
\>[27]{}\Conid{Exp}^\bullet{}\<[33]%
\>[33]{}\mid {}\<[33E]%
\>[36]{}\Conid{Prj}^\bullet\;{}\<[42]%
\>[42]{}\Conid{Var}\;{}\<[47]%
\>[47]{}\Conid{Var}\;\Conid{Exp}^\bullet{}\<[E]%
\ColumnHook
\end{hscode}\resethooks

Thanks to this update, type checking is a straightforward structural
recursive definition:
\begin{hscode}\SaveRestoreHook
\column{B}{@{}>{\hspre}l<{\hspost}@{}}%
\column{7}{@{}>{\hspre}l<{\hspost}@{}}%
\column{14}{@{}>{\hspre}l<{\hspost}@{}}%
\column{21}{@{}>{\hspre}l<{\hspost}@{}}%
\column{22}{@{}>{\hspre}l<{\hspost}@{}}%
\column{25}{@{}>{\hspre}l<{\hspost}@{}}%
\column{28}{@{}>{\hspre}l<{\hspost}@{}}%
\column{29}{@{}>{\hspre}l<{\hspost}@{}}%
\column{36}{@{}>{\hspre}l<{\hspost}@{}}%
\column{46}{@{}>{\hspre}c<{\hspost}@{}}%
\column{46E}{@{}l@{}}%
\column{47}{@{}>{\hspre}c<{\hspost}@{}}%
\column{47E}{@{}l@{}}%
\column{49}{@{}>{\hspre}l<{\hspost}@{}}%
\column{50}{@{}>{\hspre}l<{\hspost}@{}}%
\column{56}{@{}>{\hspre}l<{\hspost}@{}}%
\column{59}{@{}>{\hspre}l<{\hspost}@{}}%
\column{61}{@{}>{\hspre}l<{\hspost}@{}}%
\column{66}{@{}>{\hspre}l<{\hspost}@{}}%
\column{75}{@{}>{\hspre}l<{\hspost}@{}}%
\column{77}{@{}>{\hspre}l<{\hspost}@{}}%
\column{86}{@{}>{\hspre}l<{\hspost}@{}}%
\column{89}{@{}>{\hspre}l<{\hspost}@{}}%
\column{96}{@{}>{\hspre}l<{\hspost}@{}}%
\column{107}{@{}>{\hspre}l<{\hspost}@{}}%
\column{E}{@{}>{\hspre}l<{\hspost}@{}}%
\>[B]{}\mathbf{deriving}\;\mathbf{instance}\;\Conid{Eq}\;\Conid{Typ}{}\<[E]%
\\[\blanklineskip]%
\>[B]{}\Varid{chkE}\mathbin{::}\Conid{Exp}^\bullet\to \Conid{TypEnv}\to \Conid{Typ}\to \Conid{Bool}{}\<[E]%
\\
\>[B]{}\Varid{chkE}\;{}\<[7]%
\>[7]{}(\Conid{Lit}^\bullet\;{}\<[22]%
\>[22]{}\anonymous )\;{}\<[29]%
\>[29]{}\anonymous \;{}\<[36]%
\>[36]{}\Conid{Int}{}\<[47]%
\>[47]{}\mathrel{=}{}\<[47E]%
\>[50]{}\Conid{True}{}\<[E]%
\\
\>[B]{}\Varid{chkE}\;{}\<[7]%
\>[7]{}(\Conid{Var}^\bullet\;{}\<[22]%
\>[22]{}\Varid{x})\;{}\<[29]%
\>[29]{}\Gamma\;{}\<[36]%
\>[36]{}\Varid{c}{}\<[47]%
\>[47]{}\mathrel{=}{}\<[47E]%
\>[50]{}\Varid{maybe}\;\Conid{False}\;(\equiv \Varid{c})\;(\Varid{lookup}\;\Varid{x}\;\Gamma){}\<[E]%
\\
\>[B]{}\Varid{chkE}\;{}\<[7]%
\>[7]{}(\Conid{Typ}^\bullet\;{}\<[22]%
\>[22]{}\Varid{m}\;{}\<[25]%
\>[25]{}\Varid{a})\;{}\<[29]%
\>[29]{}\Gamma\;{}\<[36]%
\>[36]{}\Varid{c}{}\<[47]%
\>[47]{}\mathrel{=}{}\<[47E]%
\>[50]{}\Varid{a}\equiv \Varid{c}\mathrel{\wedge}{}\<[61]%
\>[61]{}\Varid{chkE}\;\Varid{m}\;\Gamma\;\Varid{c}{}\<[E]%
\\
\>[B]{}\Varid{chkE}\;{}\<[7]%
\>[7]{}(\Conid{Abs}^\bullet\;{}\<[22]%
\>[22]{}\Varid{x}\;{}\<[25]%
\>[25]{}\Varid{n})\;{}\<[29]%
\>[29]{}\Gamma\;{}\<[36]%
\>[36]{}(\Varid{a}:\!\rightarrow\Varid{b}){}\<[47]%
\>[47]{}\mathrel{=}{}\<[47E]%
\>[50]{}\Varid{chkE}\;{}\<[56]%
\>[56]{}\Varid{n}\;((\Varid{x},\Varid{a})\mathbin{:}\Gamma)\;{}\<[75]%
\>[75]{}\Varid{b}{}\<[E]%
\\
\>[B]{}\Varid{chkE}\;{}\<[7]%
\>[7]{}(\Conid{App}^\bullet\;{}\<[14]%
\>[14]{}\Varid{a}\;{}\<[22]%
\>[22]{}\Varid{l}\;{}\<[25]%
\>[25]{}\Varid{m})\;{}\<[29]%
\>[29]{}\Gamma\;{}\<[36]%
\>[36]{}\Varid{c}{}\<[47]%
\>[47]{}\mathrel{=}{}\<[47E]%
\>[50]{}\Varid{chkE}\;{}\<[56]%
\>[56]{}\Varid{l}\;{}\<[59]%
\>[59]{}\Gamma\;{}\<[66]%
\>[66]{}(\Varid{a}:\!\rightarrow\Varid{c}){}\<[77]%
\>[77]{}\mathrel{\wedge}\Varid{chkE}\;{}\<[86]%
\>[86]{}\Varid{m}\;{}\<[89]%
\>[89]{}\Gamma\;{}\<[96]%
\>[96]{}\Varid{a}{}\<[E]%
\\
\>[B]{}\Varid{chkE}\;{}\<[7]%
\>[7]{}(\Conid{Tup}^\bullet\;{}\<[22]%
\>[22]{}\Varid{m}\;{}\<[25]%
\>[25]{}\Varid{n})\;{}\<[29]%
\>[29]{}\Gamma\;{}\<[36]%
\>[36]{}(\Varid{a}:\!\!*\!\!:\Varid{b}){}\<[47]%
\>[47]{}\mathrel{=}{}\<[47E]%
\>[50]{}\Varid{chkE}\;{}\<[56]%
\>[56]{}\Varid{m}\;{}\<[59]%
\>[59]{}\Gamma\;{}\<[66]%
\>[66]{}\Varid{a}{}\<[77]%
\>[77]{}\mathrel{\wedge}\Varid{chkE}\;{}\<[86]%
\>[86]{}\Varid{n}\;{}\<[89]%
\>[89]{}\Gamma\;{}\<[96]%
\>[96]{}\Varid{b}{}\<[E]%
\\
\>[B]{}\Varid{chkE}\;{}\<[7]%
\>[7]{}(\Conid{Let}^\bullet\;{}\<[14]%
\>[14]{}\Delta\;{}\<[22]%
\>[22]{}\Varid{d}\;{}\<[25]%
\>[25]{}\Varid{n})\;{}\<[29]%
\>[29]{}\Gamma\;{}\<[36]%
\>[36]{}\Varid{c}{}\<[47]%
\>[47]{}\mathrel{=}{}\<[47E]%
\>[50]{}\Varid{chkD}\;{}\<[56]%
\>[56]{}\Varid{d}\;{}\<[59]%
\>[59]{}\Gamma\;{}\<[66]%
\>[66]{}\Delta{}\<[77]%
\>[77]{}\mathrel{\wedge}\Varid{chkE}\;{}\<[86]%
\>[86]{}\Varid{n}\;{}\<[89]%
\>[89]{}(\Delta\plus \Gamma)\;{}\<[107]%
\>[107]{}\Varid{c}{}\<[E]%
\\
\>[B]{}\Varid{chkE}\;{}\<[7]%
\>[7]{}\anonymous \;{}\<[29]%
\>[29]{}\anonymous \;{}\<[36]%
\>[36]{}\anonymous {}\<[47]%
\>[47]{}\mathrel{=}{}\<[47E]%
\>[50]{}\Conid{False}{}\<[E]%
\\[\blanklineskip]%
\>[B]{}\Varid{chkD}\mathbin{::}\Conid{Dec}^\bullet\to \Conid{TypEnv}\to \Conid{TypEnv}\to \Conid{Bool}{}\<[E]%
\\
\>[B]{}\Varid{chkD}\;{}\<[7]%
\>[7]{}(\Conid{Val}^\bullet\;\Varid{x}\;\Varid{m})\;{}\<[21]%
\>[21]{}\Gamma\;{}\<[28]%
\>[28]{}[\mskip1.5mu (\Varid{x'},\Varid{a})\mskip1.5mu]{}\<[46]%
\>[46]{}\mathrel{=}{}\<[46E]%
\>[49]{}\Varid{x}\equiv \Varid{x'}\mathrel{\wedge}\Varid{chkE}\;\Varid{m}\;\Gamma\;\Varid{a}{}\<[E]%
\\
\>[B]{}\Varid{chkD}\;{}\<[7]%
\>[7]{}(\Conid{Prj}^\bullet\;\Varid{x}\;\Varid{y}\;\Varid{l})\;{}\<[21]%
\>[21]{}\Gamma\;{}\<[28]%
\>[28]{}[\mskip1.5mu (\Varid{x'},\Varid{a}),(\Varid{y'},\Varid{b})\mskip1.5mu]{}\<[46]%
\>[46]{}\mathrel{=}{}\<[46E]%
\>[49]{}\Varid{x}\equiv \Varid{x'}\mathrel{\wedge}\Varid{y}\equiv \Varid{y'}\mathrel{\wedge}\Varid{chkE}\;\Varid{l}\;\Gamma\;(\Varid{a}:\!\!*\!\!:\Varid{b}){}\<[E]%
\\
\>[B]{}\Varid{chkD}\;{}\<[7]%
\>[7]{}\anonymous \;{}\<[21]%
\>[21]{}\anonymous \;{}\<[28]%
\>[28]{}\anonymous {}\<[46]%
\>[46]{}\mathrel{=}{}\<[46E]%
\>[49]{}\Conid{False}{}\<[E]%
\ColumnHook
\end{hscode}\resethooks

To use the printer defined earlier with the new decorated trees, the
definition should be updated so that it ignores the additional information:

\begin{hscode}\SaveRestoreHook
\column{B}{@{}>{\hspre}l<{\hspost}@{}}%
\column{10}{@{}>{\hspre}l<{\hspost}@{}}%
\column{17}{@{}>{\hspre}l<{\hspost}@{}}%
\column{42}{@{}>{\hspre}l<{\hspost}@{}}%
\column{45}{@{}>{\hspre}l<{\hspost}@{}}%
\column{49}{@{}>{\hspre}c<{\hspost}@{}}%
\column{49E}{@{}l@{}}%
\column{52}{@{}>{\hspre}c<{\hspost}@{}}%
\column{52E}{@{}l@{}}%
\column{E}{@{}>{\hspre}l<{\hspost}@{}}%
\>[B]{}\Varid{printE}^\bullet\mathbin{::}\Conid{Exp}^\bullet\to \Conid{String}{}\<[E]%
\\
\>[B]{}\Varid{printE}^\bullet\;{}\<[10]%
\>[10]{}(\Conid{App}^\bullet\;{}\<[17]%
\>[17]{}\hlt{\anonymous \;\!}\;{}\<[42]%
\>[42]{}\Varid{l}\;{}\<[45]%
\>[45]{}\Varid{m}){}\<[49]%
\>[49]{}\mathrel{=}{}\<[49E]%
\>[52]{}\ldots{}\<[52E]%
\\
\>[B]{}\Varid{printE}^\bullet\;{}\<[10]%
\>[10]{}(\Conid{Let}^\bullet\;{}\<[17]%
\>[17]{}\hlt{\anonymous \;\!}\;{}\<[42]%
\>[42]{}\Varid{d}\;{}\<[45]%
\>[45]{}\Varid{n}){}\<[49]%
\>[49]{}\mathrel{=}{}\<[49E]%
\>[52]{}\ldots{}\<[52E]%
\\
\>[B]{}\Varid{printE}^\bullet{}\<[10]%
\>[10]{}\ldots{}\<[49]%
\>[49]{}\mathrel{=}{}\<[49E]%
\>[52]{}\ldots{}\<[52E]%
\\[\blanklineskip]%
\>[B]{}\Varid{printD}^\bullet\mathbin{::}\Conid{Dec}^\bullet\to \Conid{String}{}\<[E]%
\\
\>[B]{}\Varid{printD}^\bullet{}\<[10]%
\>[10]{}\ldots{}\<[49]%
\>[49]{}\mathrel{=}{}\<[49E]%
\>[52]{}\ldots{}\<[52E]%
\ColumnHook
\end{hscode}\resethooks

To recap, so far we have defined
\begin{enumerate}[leftmargin=+.5in]
\setlength\itemsep{0em}
\item the non-decorated trees using \ensuremath{\Conid{Exp}} and \ensuremath{\Conid{Dec}} (and \ensuremath{\Conid{Typ}});
\item the decorated  trees using \ensuremath{\Conid{Exp}^\bullet} and \ensuremath{\Conid{Dec}^\bullet};
\item the printer for the non-decorated trees using \ensuremath{\Varid{printE}} and \ensuremath{\Varid{printD}}\\
      (and \ensuremath{\Varid{printT}});
\item the printer for the decorated trees using \ensuremath{\Varid{printE}^\bullet} and \ensuremath{\Varid{printD}^\bullet}; and
\item the type checker for the decorated trees using \ensuremath{\Varid{chkE}} and \ensuremath{\Varid{chkD}}.
\end{enumerate}

What about the set of functions in the parser generating
these trees, or the type inference engine applied before type checking? What
should be their types?

Common practice is either to use the decorated variants (e.g., \ensuremath{\Conid{Exp}^\bullet}) when the
additional information in decorations is needed, and use the non-decorated
variants (e.g., \ensuremath{\Conid{Exp}}) otherwise (when possible); or, use the decorated variants
everywhere.
For instance, with the former setting, they will be

\begin{hscode}\SaveRestoreHook
\column{B}{@{}>{\hspre}l<{\hspost}@{}}%
\column{E}{@{}>{\hspre}l<{\hspost}@{}}%
\>[B]{}\Varid{parseE}\mathbin{::}\Conid{String}\to \Conid{Maybe}\;\Conid{Exp}{}\<[E]%
\\
\>[B]{}\Varid{parseD}\mathbin{::}\Conid{String}\to \Conid{Maybe}\;\Conid{Dec}{}\<[E]%
\\[\blanklineskip]%
\>[B]{}\Varid{inferE}\mathbin{::}\Conid{Exp}\to \Conid{Exp}^\bullet{}\<[E]%
\\
\>[B]{}\Varid{inferD}\mathbin{::}\Conid{Dec}\to \Conid{Dec}^\bullet{}\<[E]%
\ColumnHook
\end{hscode}\resethooks

and with the later setting, they will be

\begin{hscode}\SaveRestoreHook
\column{B}{@{}>{\hspre}l<{\hspost}@{}}%
\column{E}{@{}>{\hspre}l<{\hspost}@{}}%
\>[B]{}\Varid{parseE}^\bullet\mathbin{::}\Conid{String}\to \Conid{Maybe}\;\Conid{Exp}^\bullet{}\<[E]%
\\
\>[B]{}\Varid{parseD}^\bullet\mathbin{::}\Conid{String}\to \Conid{Maybe}\;\Conid{Dec}^\bullet{}\<[E]%
\\[\blanklineskip]%
\>[B]{}\Varid{inferE}^\bullet\mathbin{::}\Conid{Exp}^\bullet\to \Conid{Exp}^\bullet{}\<[E]%
\\
\>[B]{}\Varid{inferD}^\bullet\mathbin{::}\Conid{Dec}^\bullet\to \Conid{Dec}^\bullet{}\<[E]%
\ColumnHook
\end{hscode}\resethooks

The former leads to duplication, both in the definition of data types and in the
functions defined over them.
While for the latter, there is no need for declarations of \ensuremath{\Conid{Exp}} and \ensuremath{\Conid{Dec}}, and
the corresponding set of functions whose equivalents are available for \ensuremath{\Conid{Exp}^\bullet}
(e.g., \ensuremath{\Varid{printE}} and \ensuremath{\Varid{printD}}).

On the other hand, for the latter, all functions using the decorated trees (the
only available variant then) should deal with the decorations explicitly, even
if their functionality is entirely independent of the decorations. This
entanglement is harmful: the more passes processing the trees, the larger the
set of unnecessary decorations. Furthermore, the decorations introduce
unnecessary dependencies between parts that define the decorations and parts
that are forced to depend on them since decorations are baked into their input
trees (and they do not use them). Notice, not every function can ignore the
unnecessary annotations, like \ensuremath{\Varid{printE}^\bullet} could; some, specially tree-to-tree
transformations, have to push the decorations around without looking at them, or
even worst, to generate dummy decorations to be able to apply a data constructor
with decorations.

\subsection{Tree-Decoration Solution}

As mentioned earlier, we suggest declaring ASTs with extensible data
types, and view decorations in trees as sets of extensions to the data type
declarations.

To avoid the need for explaining the details of encodings of extensible data
types prematurely, we explain our solution using an idealised syntax (i.e., a
macro) that we developed for Haskell. This allows our solution to be
independent of the implementation details (e.g., the encodings).
This syntax allows us to declare extensible data types, by labeling normal
algebraic data type declarations as extensible, and allows us to define
extensions to an extensible algebraic data type declaration by specifying\\
\indent(a) new fields to the existing data constructors of the extensible data type,\\
\indent(b) new data constructors to the extensible data type, and\\
\indent(c) new type parameters (with alpha renaming of the existing ones, if needed).

As an example, consider the definition of \ensuremath{\Conid{Exp}} from earlier, without bits
related to tuples:
\begin{hscode}\SaveRestoreHook
\column{B}{@{}>{\hspre}l<{\hspost}@{}}%
\column{8}{@{}>{\hspre}l<{\hspost}@{}}%
\column{15}{@{}>{\hspre}l<{\hspost}@{}}%
\column{20}{@{}>{\hspre}c<{\hspost}@{}}%
\column{20E}{@{}l@{}}%
\column{25}{@{}>{\hspre}l<{\hspost}@{}}%
\column{E}{@{}>{\hspre}l<{\hspost}@{}}%
\>[B]{}\Varid{i}\;{}\<[8]%
\>[8]{}\in\;{}\<[15]%
\>[15]{}\text{integers}{}\<[E]%
\\
\>[B]{}\Varid{x},\Varid{y}\;{}\<[8]%
\>[8]{}\in\;{}\<[15]%
\>[15]{}\text{variables}{}\<[E]%
\\
\>[B]{}\Conid{A},\Conid{B},\Conid{C}\;{}\<[8]%
\>[8]{}\in\;{}\<[15]%
\>[15]{}\text{Typ}{}\<[20]%
\>[20]{}\mathbin{::=}{}\<[20E]%
\>[25]{}\textbf{Int}\mid \Conid{A}\boldsymbol{\rightarrow}\Conid{B}{}\<[E]%
\\
\>[B]{}\Conid{L},\Conid{M},\Conid{N}\;{}\<[8]%
\>[8]{}\in\;{}\<[15]%
\>[15]{}\text{Exp}{}\<[20]%
\>[20]{}\mathbin{::=}{}\<[20E]%
\>[25]{}\Varid{i}\mid \Varid{x}\mid \Conid{M}\!\boldsymbol{::}\!\Conid{A}\mid \boldsymbol{\lambda}\Varid{x}\boldsymbol{.}\,\Conid{N}\mid \Conid{L}\;\Conid{M}\mid \textbf{let}\;\Conid{D}\;\textbf{in}\;\Conid{N}{}\<[E]%
\\
\>[B]{}\Conid{D}\;{}\<[8]%
\>[8]{}\in\;{}\<[15]%
\>[15]{}\text{Dec}{}\<[20]%
\>[20]{}\mathbin{::=}{}\<[20E]%
\>[25]{}\Varid{x}\,\boldsymbol{:=}\,\Conid{M}{}\<[E]%
\ColumnHook
\end{hscode}\resethooks

An extensible declaration of the above language is as follows.

\nonewlinecode
\noindent
\begin{minipage}[t]{0.5\textwidth}
\begin{hscode}\SaveRestoreHook
\column{B}{@{}>{\hspre}l<{\hspost}@{}}%
\column{3}{@{}>{\hspre}c<{\hspost}@{}}%
\column{3E}{@{}l@{}}%
\column{6}{@{}>{\hspre}l<{\hspost}@{}}%
\column{12}{@{}>{\hspre}l<{\hspost}@{}}%
\column{18}{@{}>{\hspre}l<{\hspost}@{}}%
\column{E}{@{}>{\hspre}l<{\hspost}@{}}%
\>[B]{}\mathbf{extensible}\;\mathbf{data}\;\Conid{Exp}_{\!X}{}\<[E]%
\\
\>[B]{}\hsindent{3}{}\<[3]%
\>[3]{}\mathrel{=}{}\<[3E]%
\>[6]{}\Conid{Lit}_{\!X}\;{}\<[12]%
\>[12]{}\Conid{Integer}{}\<[E]%
\\
\>[B]{}\hsindent{3}{}\<[3]%
\>[3]{}\mid {}\<[3E]%
\>[6]{}\Conid{Var}_{\!X}\;{}\<[12]%
\>[12]{}\Conid{Var}{}\<[E]%
\\
\>[B]{}\hsindent{3}{}\<[3]%
\>[3]{}\mid {}\<[3E]%
\>[6]{}\Conid{Typ}_{\!X}\;{}\<[12]%
\>[12]{}\Conid{Exp}_{\!X}\;{}\<[18]%
\>[18]{}\Conid{Typ}_{\!X}{}\<[E]%
\\
\>[B]{}\hsindent{3}{}\<[3]%
\>[3]{}\mid {}\<[3E]%
\>[6]{}\Conid{Abs}_{\!X}\;{}\<[12]%
\>[12]{}\Conid{Var}\;{}\<[18]%
\>[18]{}\Conid{Exp}_{\!X}{}\<[E]%
\\
\>[B]{}\hsindent{3}{}\<[3]%
\>[3]{}\mid {}\<[3E]%
\>[6]{}\Conid{App}_{\!X}\;{}\<[12]%
\>[12]{}\Conid{Exp}_{\!X}\;{}\<[18]%
\>[18]{}\Conid{Exp}_{\!X}{}\<[E]%
\\
\>[B]{}\hsindent{3}{}\<[3]%
\>[3]{}\mid {}\<[3E]%
\>[6]{}\Conid{Let}_{\!X}\;{}\<[12]%
\>[12]{}\Conid{Dec}_{\!X}\;{}\<[18]%
\>[18]{}\Conid{Exp}_{\!X}{}\<[E]%
\\
\>[B]{}\newline{}\<[E]%
\ColumnHook
\end{hscode}\resethooks
\end{minipage}
\begin{minipage}[t]{0.5\textwidth}
\begin{hscode}\SaveRestoreHook
\column{B}{@{}>{\hspre}l<{\hspost}@{}}%
\column{3}{@{}>{\hspre}c<{\hspost}@{}}%
\column{3E}{@{}l@{}}%
\column{6}{@{}>{\hspre}l<{\hspost}@{}}%
\column{12}{@{}>{\hspre}l<{\hspost}@{}}%
\column{17}{@{}>{\hspre}l<{\hspost}@{}}%
\column{E}{@{}>{\hspre}l<{\hspost}@{}}%
\>[B]{}\mathbf{extensible}\;\mathbf{data}\;\Conid{Typ}_{\!X}{}\<[E]%
\\
\>[B]{}\hsindent{3}{}\<[3]%
\>[3]{}\mathrel{=}{}\<[3E]%
\>[6]{}\Conid{Int}_{\!X}{}\<[E]%
\\
\>[B]{}\hsindent{3}{}\<[3]%
\>[3]{}\mid {}\<[3E]%
\>[6]{}\Conid{Typ}_{\!X}\overset{x}{:\!\rightarrow}\Conid{Typ}_{\!X}{}\<[E]%
\\[\blanklineskip]%
\>[B]{}\mathbf{extensible}\;\mathbf{data}\;\Conid{Dec}_{\!X}{}\<[E]%
\\
\>[B]{}\hsindent{3}{}\<[3]%
\>[3]{}\mathrel{=}{}\<[3E]%
\>[6]{}\Conid{Val}_{\!X}\;{}\<[12]%
\>[12]{}\Conid{Var}\;{}\<[17]%
\>[17]{}\Conid{Exp}_{\!X}{}\<[E]%
\ColumnHook
\end{hscode}\resethooks
\end{minipage}

To define a datatype as extensible, we just add the label \ensuremath{\mathbf{extensible}}
before a normal ADT declaration.

Defining printer for the extensible data type \ensuremath{\Conid{Exp}_{\!X}} above is similar to the
one for \ensuremath{\Conid{Exp}^\bullet} in that it ignores the decorations. However, in \ensuremath{\Conid{Exp}_{\!X}},
decorations are not bound to be of a specific type: they ara polymorphic,
instantiated explicitly by the extensions.

\nonewlinecode
\noindent
\begin{minipage}[t]{0.5\textwidth}
\begin{hscode}\SaveRestoreHook
\column{B}{@{}>{\hspre}l<{\hspost}@{}}%
\column{10}{@{}>{\hspre}l<{\hspost}@{}}%
\column{17}{@{}>{\hspre}l<{\hspost}@{}}%
\column{20}{@{}>{\hspre}l<{\hspost}@{}}%
\column{23}{@{}>{\hspre}l<{\hspost}@{}}%
\column{30}{@{}>{\hspre}l<{\hspost}@{}}%
\column{34}{@{}>{\hspre}c<{\hspost}@{}}%
\column{34E}{@{}l@{}}%
\column{37}{@{}>{\hspre}l<{\hspost}@{}}%
\column{E}{@{}>{\hspre}l<{\hspost}@{}}%
\>[B]{}\Varid{printE}_{\!\!X}\;{}\<[10]%
\>[10]{}(\Conid{Lit}_{\!X}\;{}\<[17]%
\>[17]{}\Varid{i}\;{}\<[23]%
\>[23]{}\!\!\boldsymbol{\oplus}\!\!\;{}\<[30]%
\>[30]{}\anonymous ){}\<[34]%
\>[34]{}\mathrel{=}{}\<[34E]%
\>[37]{}\ldots{}\<[E]%
\\
\>[B]{}\Varid{printE}_{\!\!X}\;{}\<[10]%
\>[10]{}(\Conid{Var}_{\!X}\;{}\<[17]%
\>[17]{}\Varid{x}\;{}\<[23]%
\>[23]{}\!\!\boldsymbol{\oplus}\!\!\;{}\<[30]%
\>[30]{}\anonymous ){}\<[34]%
\>[34]{}\mathrel{=}{}\<[34E]%
\>[37]{}\ldots{}\<[E]%
\\
\>[B]{}\Varid{printE}_{\!\!X}\;{}\<[10]%
\>[10]{}(\Conid{Typ}_{\!X}\;{}\<[17]%
\>[17]{}\Varid{m}\;{}\<[20]%
\>[20]{}\Varid{a}\;{}\<[23]%
\>[23]{}\!\!\boldsymbol{\oplus}\!\!\;{}\<[30]%
\>[30]{}\anonymous ){}\<[34]%
\>[34]{}\mathrel{=}{}\<[34E]%
\>[37]{}\ldots{}\<[E]%
\\
\>[B]{}\Varid{printE}_{\!\!X}\;{}\<[10]%
\>[10]{}(\Conid{Abs}_{\!X}\;{}\<[17]%
\>[17]{}\Varid{x}\;{}\<[20]%
\>[20]{}\Varid{n}\;{}\<[23]%
\>[23]{}\!\!\boldsymbol{\oplus}\!\!\;{}\<[30]%
\>[30]{}\anonymous ){}\<[34]%
\>[34]{}\mathrel{=}{}\<[34E]%
\>[37]{}\ldots{}\<[E]%
\\
\>[B]{}\Varid{printE}_{\!\!X}\;{}\<[10]%
\>[10]{}(\Conid{App}_{\!X}\;{}\<[17]%
\>[17]{}\Varid{l}\;{}\<[20]%
\>[20]{}\Varid{m}\;{}\<[23]%
\>[23]{}\!\!\boldsymbol{\oplus}\!\!\;{}\<[30]%
\>[30]{}\anonymous ){}\<[34]%
\>[34]{}\mathrel{=}{}\<[34E]%
\>[37]{}\ldots{}\<[E]%
\\
\>[B]{}\Varid{printE}_{\!\!X}\;{}\<[10]%
\>[10]{}(\Conid{Let}_{\!X}\;{}\<[17]%
\>[17]{}\Varid{d}\;{}\<[20]%
\>[20]{}\Varid{n}\;{}\<[23]%
\>[23]{}\!\!\boldsymbol{\oplus}\!\!\;{}\<[30]%
\>[30]{}\anonymous ){}\<[34]%
\>[34]{}\mathrel{=}{}\<[34E]%
\>[37]{}\ldots{}\<[E]%
\\
\>[B]{}\Varid{printE}_{\!\!X}\;{}\<[10]%
\>[10]{}(\Conid{Exp}_{\!X}\;{}\<[23]%
\>[23]{}\!\!\boldsymbol{\oplus}\!\!\;{}\<[30]%
\>[30]{}\Varid{e}){}\<[34]%
\>[34]{}\mathrel{=}{}\<[34E]%
\>[37]{}?\Varid{printE}_{\!\text{Ext}}\;\Varid{e}{}\<[E]%
\\
\>[B]{}\newline{}\<[E]%
\ColumnHook
\end{hscode}\resethooks
\end{minipage}
\begin{minipage}[t]{0.5\textwidth}
\begin{hscode}\SaveRestoreHook
\column{B}{@{}>{\hspre}l<{\hspost}@{}}%
\column{10}{@{}>{\hspre}l<{\hspost}@{}}%
\column{21}{@{}>{\hspre}l<{\hspost}@{}}%
\column{28}{@{}>{\hspre}l<{\hspost}@{}}%
\column{32}{@{}>{\hspre}c<{\hspost}@{}}%
\column{32E}{@{}l@{}}%
\column{35}{@{}>{\hspre}l<{\hspost}@{}}%
\column{47}{@{}>{\hspre}l<{\hspost}@{}}%
\column{E}{@{}>{\hspre}l<{\hspost}@{}}%
\>[B]{}\Varid{printT}_{\!\!X}\;{}\<[10]%
\>[10]{}(\Conid{Int}_{\!X}\;{}\<[21]%
\>[21]{}\!\!\boldsymbol{\oplus}\!\!\;{}\<[28]%
\>[28]{}\anonymous ){}\<[32]%
\>[32]{}\mathrel{=}{}\<[32E]%
\>[35]{}\ldots{}\<[E]%
\\
\>[B]{}\Varid{printT}_{\!\!X}\;{}\<[10]%
\>[10]{}(\Varid{a}\overset{x}{:\!\rightarrow}\Varid{b}\;{}\<[21]%
\>[21]{}\!\!\boldsymbol{\oplus}\!\!\;{}\<[28]%
\>[28]{}\anonymous ){}\<[32]%
\>[32]{}\mathrel{=}{}\<[32E]%
\>[35]{}\ldots{}\<[E]%
\\
\>[B]{}\Varid{printT}_{\!\!X}\;{}\<[10]%
\>[10]{}(\Conid{Typ}_{\!X}\;{}\<[21]%
\>[21]{}\!\!\boldsymbol{\oplus}\!\!\;{}\<[28]%
\>[28]{}\Varid{e}){}\<[32]%
\>[32]{}\mathrel{=}{}\<[32E]%
\>[35]{}?\Varid{printT}_{\!\text{Ext}}\;{}\<[47]%
\>[47]{}\Varid{e}{}\<[E]%
\\[\blanklineskip]%
\>[B]{}\Varid{printD}_{\!\!X}\;{}\<[10]%
\>[10]{}(\Conid{Val}_{\!X}\;\Varid{x}\;\Varid{m}\;{}\<[21]%
\>[21]{}\!\!\boldsymbol{\oplus}\!\!\;{}\<[28]%
\>[28]{}\anonymous ){}\<[32]%
\>[32]{}\mathrel{=}{}\<[32E]%
\>[35]{}\ldots{}\<[E]%
\\
\>[B]{}\Varid{printD}_{\!\!X}\;{}\<[10]%
\>[10]{}(\Conid{Dec}_{\!X}\;{}\<[21]%
\>[21]{}\!\!\boldsymbol{\oplus}\!\!\;{}\<[28]%
\>[28]{}\Varid{e}){}\<[32]%
\>[32]{}\mathrel{=}{}\<[32E]%
\>[35]{}?\Varid{printD}_{\!\text{Ext}}\;{}\<[47]%
\>[47]{}\Varid{e}{}\<[E]%
\ColumnHook
\end{hscode}\resethooks
\end{minipage}

The pattern syntax \ensuremath{\Conid{C}_{\Conid{X}}\;\Conid{P}_{\mathrm{1}}\ldots\Conid{P}_{\Varid{n}}\;\!\!\boldsymbol{\oplus}\!\!\;\Conid{P'}} describes matching on
the existing fields of the the constructor \ensuremath{\Conid{C}_{\Conid{X}}} of an extensible datatype by
patterns \ensuremath{\Conid{P}_{\mathrm{1}}} to \ensuremath{\Conid{P}_{\Varid{n}}} and matching on the new field (introducesd via
extensions) to that constructor by patterns \ensuremath{\Conid{P'}}. The pattern syntax
\ensuremath{\Conid{P'}\;\!\!\boldsymbol{\oplus}\!\!\;\Conid{T}_{\Conid{X}}} describes matching on the new constructors (introduced via
extensions) to the extensible data type \ensuremath{\Conid{T}_{\Conid{X}}} by pattern \ensuremath{\Conid{P'}}.

In the above example, to avoid cluttering the presentation with explicit passing
of functions, we have used implicit parameters to introduce functions that are
applied to the extensions. We could as well use methods of a type class (say
named \ensuremath{\Conid{Printable}}), or normal explicit passing of functions.

We can use a similar syntax to construct values of an extensible data type. For
instance, consider the following.

\begin{hscode}\SaveRestoreHook
\column{B}{@{}>{\hspre}l<{\hspost}@{}}%
\column{E}{@{}>{\hspre}l<{\hspost}@{}}%
\>[B]{}\mathbf{data}\;\Conid{SrcSpan}\mathrel{=}\Conid{SrcSpan}\;\{\mskip1.5mu \Varid{begins}\mathbin{::}\Conid{Int},\Varid{ends}\mathbin{::}\Conid{Int}\mskip1.5mu\}{}\<[E]%
\\[\blanklineskip]%
\>[B]{}\Varid{myLit}\mathrel{=}\Conid{Lit}_{\!X}\;\mathrm{42}\;\!\!\boldsymbol{\oplus}\!\!\;(\Conid{SrcSpan}\;\mathrm{0}\;\mathrm{2}){}\<[E]%
\ColumnHook
\end{hscode}\resethooks

It is meant to construct an integer literal value of \ensuremath{\Conid{Exp}_{\!X}}, with an extension
field describing its position in a source file. Looking at its infered type is
instructional: the infered type is \ensuremath{(\xi\;\text{\tt \char34 LitX\char34}\mathord{\sim}\Conid{SrcSpan})\Rightarrow \Conid{Exp}_{\!X}\;\!\!\boldsymbol{\oplus}\!\!\;\xi}.
The inferred type reads as ``there is an extension named \ensuremath{\xi} to \ensuremath{\Conid{Exp}_{\!X}} that for
the constructor \ensuremath{\Conid{Lit}_{\!X}} (i.e., the new field introduced by the extension \ensuremath{\xi} in
\ensuremath{\Conid{Lit}_{\!X}}) it is of the type \ensuremath{\Conid{SrcSpan}}''.

Unfortunately, a type of this form is not currently accepted by GHC: to keep the
type-level machinery consitent with the type inference in Haskell, type
functions (i.e., type families) in GHC are not first class, and they cannot be
quantified over at the type-level. This means \ensuremath{\xi} should be a concrete type
constructor. We can, for instance choose \ensuremath{\xi} to be the constant functor \ensuremath{\Conid{Const}\;\Conid{SrcSpan}}, setting the type of all new fields (and constructor) introduced by the
extension to be \ensuremath{\Conid{SrcSpan}}. That is, we can write \ensuremath{\Conid{Lit}_{\!X}\;\mathrm{42}\;\!\!\boldsymbol{\oplus}\!\!\;(\Conid{Const}\;(\Conid{SrcSpan}\;\mathrm{0}\;\mathrm{2}))} instead, whose infered type is \ensuremath{\Conid{Exp}_{\!X}\;\!\!\boldsymbol{\oplus}\!\!\;(\Conid{Const}\;\Conid{SrcSpan})}.
To have a more refined control over the type of each extension, and to avoid
problems with type inference, in what follows we describe a way to declare extensions.
Declaring extensions, as opposed to having them inferred from the context (or by
complex type annotations) is synonymous to having declared records as opposed to
annonymous records; declarated constructors as opposed to annonymous variants;
or iso-recursive types, as opposed to equi-recursive types.

Let us start with an example. The extension that introduces the bits related to
tuples that we dropped earlier (the extension form (b) above), and that also
introduces the infered type decorations discussed earlier (the extension form
(a) above) is defined as follows.

\noindent
\begin{minipage}[t]{0.5\textwidth}
\noindent
\begin{hscode}\SaveRestoreHook
\column{B}{@{}>{\hspre}l<{\hspost}@{}}%
\column{3}{@{}>{\hspre}c<{\hspost}@{}}%
\column{3E}{@{}l@{}}%
\column{7}{@{}>{\hspre}l<{\hspost}@{}}%
\column{14}{@{}>{\hspre}l<{\hspost}@{}}%
\column{23}{@{}>{\hspre}l<{\hspost}@{}}%
\column{29}{@{}>{\hspre}l<{\hspost}@{}}%
\column{33}{@{}>{\hspre}l<{\hspost}@{}}%
\column{E}{@{}>{\hspre}l<{\hspost}@{}}%
\>[B]{}\mathbf{type}\;\Conid{TypEnv}_{\!X}^\bullet\mathrel{=}[\mskip1.5mu (\Conid{Var},\Conid{Typ}_{\!X}^\bullet)\mskip1.5mu]{}\<[E]%
\\[\blanklineskip]%
\>[B]{}\mathbf{data}\;{}\<[7]%
\>[7]{}\Conid{Exp}_{\!X}^\bullet\;{}\<[14]%
\>[14]{}\mathbf{extends}\;{}\<[23]%
\>[23]{}\Conid{Exp}_{\!X}{}\<[E]%
\\
\>[B]{}\hsindent{3}{}\<[3]%
\>[3]{}\mathrel{=}{}\<[3E]%
\>[7]{}\Conid{Tup}_{\!X}^\bullet\;\Conid{Exp}_{\!X}^\bullet\;\Conid{Exp}_{\!X}^\bullet{}\<[E]%
\\
\>[B]{}\hsindent{3}{}\<[3]%
\>[3]{}\mid {}\<[3E]%
\>[7]{}\Conid{Lit}_{\!X}^\bullet\;{}\<[14]%
\>[14]{}\mathbf{extends}\;{}\<[23]%
\>[23]{}\Conid{Lit}_{\!X}\;{}\<[29]%
\>[29]{}\mathbf{by}\;{}\<[33]%
\>[33]{}\varnothing{}\<[E]%
\\
\>[B]{}\hsindent{3}{}\<[3]%
\>[3]{}\mid {}\<[3E]%
\>[7]{}\Conid{Var}_{\!X}^\bullet\;{}\<[14]%
\>[14]{}\mathbf{extends}\;{}\<[23]%
\>[23]{}\Conid{Var}_{\!X}\;{}\<[29]%
\>[29]{}\mathbf{by}\;{}\<[33]%
\>[33]{}\varnothing{}\<[E]%
\\
\>[B]{}\hsindent{3}{}\<[3]%
\>[3]{}\mid {}\<[3E]%
\>[7]{}\Conid{Typ}_{\!X}^\bullet\;{}\<[14]%
\>[14]{}\mathbf{extends}\;{}\<[23]%
\>[23]{}\Conid{Typ}_{\!X}\;{}\<[29]%
\>[29]{}\mathbf{by}\;{}\<[33]%
\>[33]{}\varnothing{}\<[E]%
\\
\>[B]{}\hsindent{3}{}\<[3]%
\>[3]{}\mid {}\<[3E]%
\>[7]{}\Conid{Abs}_{\!X}^\bullet\;{}\<[14]%
\>[14]{}\mathbf{extends}\;{}\<[23]%
\>[23]{}\Conid{Abs}_{\!X}\;{}\<[29]%
\>[29]{}\mathbf{by}\;{}\<[33]%
\>[33]{}\varnothing{}\<[E]%
\\
\>[B]{}\hsindent{3}{}\<[3]%
\>[3]{}\mid {}\<[3E]%
\>[7]{}\Conid{App}_{\!X}^\bullet\;{}\<[14]%
\>[14]{}\mathbf{extends}\;{}\<[23]%
\>[23]{}\Conid{App}_{\!X}\;{}\<[29]%
\>[29]{}\mathbf{by}\;{}\<[33]%
\>[33]{}\Conid{Typ}_{\!X}^\bullet{}\<[E]%
\\
\>[B]{}\hsindent{3}{}\<[3]%
\>[3]{}\mid {}\<[3E]%
\>[7]{}\Conid{Let}_{\!X}^\bullet\;{}\<[14]%
\>[14]{}\mathbf{extends}\;{}\<[23]%
\>[23]{}\Conid{Let}_{\!X}\;{}\<[29]%
\>[29]{}\mathbf{by}\;{}\<[33]%
\>[33]{}\Conid{TypEnv}_{\!X}^\bullet{}\<[E]%
\\[\blanklineskip]%
\>[B]{}\newline{}\<[E]%
\ColumnHook
\end{hscode}\resethooks
\end{minipage}
\begin{minipage}[t]{0.5\textwidth}
\begin{hscode}\SaveRestoreHook
\column{B}{@{}>{\hspre}l<{\hspost}@{}}%
\column{3}{@{}>{\hspre}c<{\hspost}@{}}%
\column{3E}{@{}l@{}}%
\column{7}{@{}>{\hspre}l<{\hspost}@{}}%
\column{14}{@{}>{\hspre}l<{\hspost}@{}}%
\column{16}{@{}>{\hspre}l<{\hspost}@{}}%
\column{22}{@{}>{\hspre}l<{\hspost}@{}}%
\column{25}{@{}>{\hspre}l<{\hspost}@{}}%
\column{33}{@{}>{\hspre}l<{\hspost}@{}}%
\column{37}{@{}>{\hspre}l<{\hspost}@{}}%
\column{E}{@{}>{\hspre}l<{\hspost}@{}}%
\>[B]{}\mathbf{data}\;{}\<[7]%
\>[7]{}\Conid{Typ}_{\!X}^\bullet\;\mathbf{extends}\;\Conid{Typ}_{\!X}{}\<[E]%
\\
\>[B]{}\hsindent{3}{}\<[3]%
\>[3]{}\mathrel{=}{}\<[3E]%
\>[7]{}\Conid{Typ}_{\!X}^\bullet:\!\!*\!\!:\Conid{Typ}_{\!X}^\bullet{}\<[E]%
\\
\>[B]{}\hsindent{3}{}\<[3]%
\>[3]{}\mid {}\<[3E]%
\>[7]{}\Conid{Int}_{\!X}^\bullet\;{}\<[16]%
\>[16]{}\mathbf{extends}\;{}\<[25]%
\>[25]{}\Conid{Int}_{\!X}\;{}\<[33]%
\>[33]{}\mathbf{by}\;{}\<[37]%
\>[37]{}\varnothing{}\<[E]%
\\
\>[B]{}\hsindent{3}{}\<[3]%
\>[3]{}\mid {}\<[3E]%
\>[7]{}(\overset{x}{:\!\rightarrow}^\bullet)\;{}\<[16]%
\>[16]{}\mathbf{extends}\;{}\<[25]%
\>[25]{}(\overset{x}{:\!\rightarrow})\;{}\<[33]%
\>[33]{}\mathbf{by}\;{}\<[37]%
\>[37]{}\varnothing{}\<[E]%
\\[\blanklineskip]%
\>[B]{}\mathbf{data}\;{}\<[7]%
\>[7]{}\Conid{Dec}_{\!X}^\bullet\;\mathbf{extends}\;{}\<[22]%
\>[22]{}\Conid{Dec}_{\!X}{}\<[E]%
\\
\>[B]{}\hsindent{3}{}\<[3]%
\>[3]{}\mathrel{=}{}\<[3E]%
\>[7]{}\Conid{Prj}_{\!X}^\bullet\;{}\<[14]%
\>[14]{}\Conid{Var}\;\Conid{Var}\;\Conid{Exp}_{\!X}^\bullet{}\<[E]%
\\
\>[B]{}\hsindent{3}{}\<[3]%
\>[3]{}\mid {}\<[3E]%
\>[7]{}\Conid{Val}_{\!X}^\bullet\;{}\<[16]%
\>[16]{}\mathbf{extends}\;{}\<[25]%
\>[25]{}\Conid{Val}_{\!X}\;{}\<[33]%
\>[33]{}\mathbf{by}\;{}\<[37]%
\>[37]{}\varnothing{}\<[E]%
\ColumnHook
\end{hscode}\resethooks
\end{minipage}

The syntax \ensuremath{\mathbf{data}\;\Conid{T}\;\alpha_1\ldots\alpha_n\;\mathbf{extends}\;\Conid{T}_{\Conid{X}}\;\beta_1\ldots\beta_m\mathrel{=}\ldots} declares \ensuremath{\Conid{T}} with
type variables \ensuremath{\Varid{a}_{\mathrm{1}}} to \ensuremath{\Varid{a}_{\Varid{n}}}, as an extension of \ensuremath{\Conid{T}_{\Conid{X}}} where type variables of
\ensuremath{\Conid{T}_{\Conid{X}}} in the position \ensuremath{\mathrm{1}} to \ensuremath{\Varid{m}} are set to variables \ensuremath{\Varid{b}_{\mathrm{1}}} to \ensuremath{\Varid{b}_{\Varid{m}}}. A
variables \ensuremath{\Varid{b}_{\Varid{i}}} should be one of the variables \ensuremath{\Varid{a}_{\mathrm{1}}} to \ensuremath{\Varid{a}_{\Varid{n}}} (not necessarily
the same order), so we have: \ensuremath{\Varid{b}_{\Varid{i}}\;\in\;\{\mskip1.5mu \Varid{a}_{\mathrm{1}}\ldots\Varid{a}_{\Varid{n}}\mskip1.5mu\}} and \ensuremath{\Varid{m}\leq \Varid{n}}.  The
declarations allows for extension of form (c) mentioned earlier, where an
extended data type may have an additional set of variables compared to the base
extensible data type. We refer to \ensuremath{\Conid{T}} as the extending data type, and \ensuremath{\Conid{T}_{\Conid{X}}} as
the base data type.

The syntax \ensuremath{\Conid{C}\;\mathbf{extends}\;\Conid{C}_{\Conid{X}}\;\mathbf{by}\;\Conid{T}_{\mathrm{1}}\ldots\Conid{T}_{\Varid{n}}} declares the constructor \ensuremath{\Conid{C}} for the
extending data type that has all the fields of the constructor \ensuremath{\Conid{C}_{\Conid{X}}} of the base
data type in addition to the fields of the types \ensuremath{\Conid{T}_{\mathrm{1}}} to \ensuremath{\Conid{T}_{\Varid{n}}}.

The constructor declarations without \ensuremath{\mathbf{extends}\ldots\mathbf{by}\ldots} are the new
constructors that the extending data type extends the base data type with.

The data types \ensuremath{\Conid{Typ}_{\!X}^\bullet}, \ensuremath{\Conid{Exp}_{\!X}^\bullet}, and \ensuremath{\Conid{Dec}_{\!X}^\bullet} are equivalent to \ensuremath{\Conid{Typ}^\bullet}, \ensuremath{\Conid{Exp}^\bullet},
and \ensuremath{\Conid{Dec}^\bullet} in that they have data constructors of the same signature and can
carry the same information (i.e., isomorphic).  The type checking function for
\ensuremath{\Conid{Exp}_{\!X}^\bullet} and \ensuremath{\Conid{Dec}_{\!X}^\bullet} would be exactly the same as the ones for \ensuremath{\Conid{Exp}^\bullet} and
\ensuremath{\Conid{Dec}^\bullet}. Same applies for the printer function.  Yet, thanks to extensibility,
the extending data types can reuse the one defined for the base data type.

\begin{hscode}\SaveRestoreHook
\column{B}{@{}>{\hspre}l<{\hspost}@{}}%
\column{11}{@{}>{\hspre}l<{\hspost}@{}}%
\column{13}{@{}>{\hspre}l<{\hspost}@{}}%
\column{14}{@{}>{\hspre}l<{\hspost}@{}}%
\column{19}{@{}>{\hspre}l<{\hspost}@{}}%
\column{22}{@{}>{\hspre}l<{\hspost}@{}}%
\column{36}{@{}>{\hspre}c<{\hspost}@{}}%
\column{36E}{@{}l@{}}%
\column{39}{@{}>{\hspre}l<{\hspost}@{}}%
\column{45}{@{}>{\hspre}c<{\hspost}@{}}%
\column{45E}{@{}l@{}}%
\column{48}{@{}>{\hspre}l<{\hspost}@{}}%
\column{49}{@{}>{\hspre}l<{\hspost}@{}}%
\column{59}{@{}>{\hspre}l<{\hspost}@{}}%
\column{62}{@{}>{\hspre}c<{\hspost}@{}}%
\column{62E}{@{}l@{}}%
\column{66}{@{}>{\hspre}l<{\hspost}@{}}%
\column{76}{@{}>{\hspre}l<{\hspost}@{}}%
\column{77}{@{}>{\hspre}l<{\hspost}@{}}%
\column{86}{@{}>{\hspre}l<{\hspost}@{}}%
\column{87}{@{}>{\hspre}l<{\hspost}@{}}%
\column{89}{@{}>{\hspre}l<{\hspost}@{}}%
\column{E}{@{}>{\hspre}l<{\hspost}@{}}%
\>[B]{}\Varid{printT}_{\!\text{Ext}}^\bullet\mathbin{::}\Conid{Typ}_{\!X}^\bullet\mathord{\sim}\Conid{Typ}_{\!X}\;\!\!\boldsymbol{\oplus}\!\!\;\xi\Rightarrow \xi\;\text{\tt \char34 TypX\char34}\to \Conid{String}{}\<[E]%
\\
\>[B]{}\Varid{printT}_{\!\text{Ext}}^\bullet\;{}\<[13]%
\>[13]{}(\underline{\hspace{5pt}}\!\oplus\!\!\;{}\<[22]%
\>[22]{}\Varid{a}:\!\!*\!\!:\Varid{b}){}\<[36]%
\>[36]{}\mathrel{=}{}\<[36E]%
\>[39]{}\text{\tt \char34 (\char34}{}\<[45]%
\>[45]{}\plus {}\<[45E]%
\>[49]{}\Varid{printT}_{\!\!X}^\bullet\;{}\<[59]%
\>[59]{}\Varid{a}{}\<[62]%
\>[62]{}\plus {}\<[62E]%
\>[66]{}\text{\tt \char34 )~×~\char34}\plus {}\<[77]%
\>[77]{}\Varid{printT}_{\!\!X}^\bullet\;{}\<[87]%
\>[87]{}\Varid{b}{}\<[E]%
\\[\blanklineskip]%
\>[B]{}\Varid{printE}_{\!\text{Ext}}^\bullet\mathbin{::}\Conid{Exp}_{\!X}^\bullet\mathord{\sim}\Conid{Exp}_{\!X}\;\!\!\boldsymbol{\oplus}\!\!\;\xi\Rightarrow \xi\;\text{\tt \char34 ExpX\char34}\to \Conid{String}{}\<[E]%
\\
\>[B]{}\Varid{printE}_{\!\text{Ext}}^\bullet\;{}\<[13]%
\>[13]{}(\underline{\hspace{5pt}}\!\oplus\!\!\;{}\<[22]%
\>[22]{}\Conid{Tup}_{\!X}^\bullet\;\Varid{m}\;\Varid{n}){}\<[36]%
\>[36]{}\mathrel{=}{}\<[36E]%
\>[39]{}\text{\tt \char34 (\char34}{}\<[45]%
\>[45]{}\plus {}\<[45E]%
\>[49]{}\Varid{printE}_{\!\!X}^\bullet\;{}\<[59]%
\>[59]{}\Varid{m}{}\<[62]%
\>[62]{}\plus {}\<[62E]%
\>[66]{}\text{\tt \char34 ~,~\char34}\plus {}\<[76]%
\>[76]{}\Varid{printE}_{\!\!X}^\bullet\;{}\<[86]%
\>[86]{}\Varid{n}{}\<[89]%
\>[89]{}\plus \text{\tt \char34 )\char34}{}\<[E]%
\\[\blanklineskip]%
\>[B]{}\Varid{printE}_{\!\text{Ext}}^\bullet\mathbin{::}\Conid{Dec}_{\!X}^\bullet\mathord{\sim}\Conid{Dec}_{\!X}\;\!\!\boldsymbol{\oplus}\!\!\;\xi\Rightarrow \xi\;\text{\tt \char34 DecX\char34}\to \Conid{String}{}\<[E]%
\\
\>[B]{}\Varid{printD}_{\!\text{Ext}}^\bullet\;{}\<[13]%
\>[13]{}(\underline{\hspace{5pt}}\!\oplus\!\!\;{}\<[22]%
\>[22]{}\Conid{Prj}_{\!X}^\bullet\;\Varid{x}\;\Varid{y}\;\Varid{l}){}\<[36]%
\>[36]{}\mathrel{=}{}\<[36E]%
\>[39]{}\text{\tt \char34 (\char34}\plus \Varid{x}\plus \text{\tt \char34 ~,~\char34}\plus \Varid{y}\plus \text{\tt \char34 )~:=~\char34}\plus \Varid{printE}_{\!\!X}^\bullet\;\Varid{l}{}\<[E]%
\\[\blanklineskip]%
\>[B]{}\Varid{printT}_{\!\!X}^\bullet{}\<[11]%
\>[11]{}\mathbin{::}\Conid{Typ}_{\!X}^\bullet\to \Conid{String}{}\<[E]%
\\
\>[B]{}\Varid{printT}_{\!\!X}^\bullet{}\<[11]%
\>[11]{}\mathrel{=}{}\<[14]%
\>[14]{}\mathbf{let}\;{}\<[19]%
\>[19]{}?\Varid{printT}_{\!\text{Ext}}\mathrel{=}\Varid{printT}_{\!\text{Ext}}^\bullet\;\mathbf{in}\;{}\<[48]%
\>[48]{}\Varid{printT}_{\!\!X}{}\<[E]%
\ColumnHook
\end{hscode}\resethooks

\noindent
\begin{minipage}[t]{0.5\textwidth}
\noindent
\begin{hscode}\SaveRestoreHook
\column{B}{@{}>{\hspre}l<{\hspost}@{}}%
\column{3}{@{}>{\hspre}c<{\hspost}@{}}%
\column{3E}{@{}l@{}}%
\column{6}{@{}>{\hspre}l<{\hspost}@{}}%
\column{11}{@{}>{\hspre}l<{\hspost}@{}}%
\column{E}{@{}>{\hspre}l<{\hspost}@{}}%
\>[B]{}\Varid{printE}_{\!\!X}^\bullet\mathbin{::}\Conid{Exp}_{\!X}^\bullet\to \Conid{String}{}\<[E]%
\\
\>[B]{}\Varid{printE}_{\!\!X}^\bullet{}\<[E]%
\\
\>[B]{}\hsindent{3}{}\<[3]%
\>[3]{}\mathrel{=}{}\<[3E]%
\>[6]{}\mathbf{let}\;{}\<[11]%
\>[11]{}?\Varid{printT}_{\!\text{Ext}}\mathrel{=}\Varid{printT}_{\!\text{Ext}}^\bullet{}\<[E]%
\\
\>[11]{}?\Varid{printE}_{\!\text{Ext}}\mathrel{=}\Varid{printE}_{\!\text{Ext}}^\bullet{}\<[E]%
\\
\>[11]{}?\Varid{printD}_{\!\text{Ext}}\mathrel{=}\Varid{printD}_{\!\text{Ext}}^\bullet{}\<[E]%
\\
\>[6]{}\mathbf{in}\;{}\<[11]%
\>[11]{}\Varid{printE}_{\!\!X}{}\<[E]%
\\[\blanklineskip]%
\>[B]{}\newline{}\<[E]%
\ColumnHook
\end{hscode}\resethooks
\end{minipage}
\begin{minipage}[t]{0.5\textwidth}
\begin{hscode}\SaveRestoreHook
\column{B}{@{}>{\hspre}l<{\hspost}@{}}%
\column{3}{@{}>{\hspre}c<{\hspost}@{}}%
\column{3E}{@{}l@{}}%
\column{6}{@{}>{\hspre}l<{\hspost}@{}}%
\column{11}{@{}>{\hspre}l<{\hspost}@{}}%
\column{E}{@{}>{\hspre}l<{\hspost}@{}}%
\>[B]{}\Varid{printD}_{\!\!X}^\bullet\mathbin{::}\Conid{Dec}_{\!X}^\bullet\to \Conid{String}{}\<[E]%
\\
\>[B]{}\Varid{printD}_{\!\!X}^\bullet{}\<[E]%
\\
\>[B]{}\hsindent{3}{}\<[3]%
\>[3]{}\mathrel{=}{}\<[3E]%
\>[6]{}\mathbf{let}\;{}\<[11]%
\>[11]{}?\Varid{printT}_{\!\text{Ext}}\mathrel{=}\Varid{printT}_{\!\text{Ext}}^\bullet{}\<[E]%
\\
\>[11]{}?\Varid{printE}_{\!\text{Ext}}\mathrel{=}\Varid{printE}_{\!\text{Ext}}^\bullet{}\<[E]%
\\
\>[11]{}?\Varid{printD}_{\!\text{Ext}}\mathrel{=}\Varid{printD}_{\!\text{Ext}}^\bullet{}\<[E]%
\\
\>[6]{}\mathbf{in}\;{}\<[11]%
\>[11]{}\Varid{printD}_{\!\!X}{}\<[E]%
\ColumnHook
\end{hscode}\resethooks
\end{minipage}

The pattern syntax \ensuremath{\underline{\hspace{5pt}}\!\oplus\!\!\;\Conid{C}\;\Conid{P}_{\mathrm{1}}\ldots\Conid{P}_{\Varid{n}}} describes matching only on the
extensions (e.g., the decorations) in the constructor \ensuremath{\Conid{K}} of an extending data
type by patterns \ensuremath{\Conid{P}_{\mathrm{1}}} to \ensuremath{\Conid{P}_{\Varid{n}}}. The type \ensuremath{\Conid{T}\mathord{\sim}\Conid{T}_{\Conid{X}}\;\!\!\boldsymbol{\oplus}\!\!\;\xi\Rightarrow \xi\;\text{\tt \char34 S\char34}} can be
read as ``\ensuremath{\xi} is the extension by which \ensuremath{\Conid{T}} extends \ensuremath{\Conid{T}_{\Conid{X}}}, and we are
interested in the specific extension at \ensuremath{\Conid{S}}, where \ensuremath{\Conid{S}} ranges over the names of
the constructors in \ensuremath{\Conid{T}_{\Conid{X}}} (and data types defined mutually-recursively with it)
to represent new fields to that constructor, and the name \ensuremath{\Conid{T}_{\Conid{X}}} itself (and data
types defined mutually-recursively with it) to represent the set of new
constructors . We could as well introduce the notation \ensuremath{(\Conid{T}\;\!\!\ominus\!\!\;\Conid{T}_{\Conid{X}})\;\text{\tt \char34 S\char34}} to
express the same.

Notice how we managed to define the printer in a compositional way.  However, in
general, it may not be practically possible to define functions over an
extending data type as a composition of functions defined separately over the
base data type and the pieces of extensions. Yet still, extensible data types
allow for reuse of data type declarations even if functions defined over them
cannot be reused.

Following the same techniques, we can reuse the base data type delcarations to
define data types equivalent to the ones representing the non-decorated AST,
i.e., \ensuremath{\Conid{Typ}}, \ensuremath{\Conid{Exp}}, and \ensuremath{\Conid{Dec}}:

\noindent
\begin{minipage}[t]{0.5\textwidth}
\noindent
\begin{hscode}\SaveRestoreHook
\column{B}{@{}>{\hspre}l<{\hspost}@{}}%
\column{3}{@{}>{\hspre}c<{\hspost}@{}}%
\column{3E}{@{}l@{}}%
\column{6}{@{}>{\hspre}c<{\hspost}@{}}%
\column{6E}{@{}l@{}}%
\column{7}{@{}>{\hspre}l<{\hspost}@{}}%
\column{14}{@{}>{\hspre}l<{\hspost}@{}}%
\column{23}{@{}>{\hspre}l<{\hspost}@{}}%
\column{29}{@{}>{\hspre}l<{\hspost}@{}}%
\column{33}{@{}>{\hspre}l<{\hspost}@{}}%
\column{E}{@{}>{\hspre}l<{\hspost}@{}}%
\>[B]{}\mathbf{type}\;\Conid{TypEnv}_{\!X}^\circ\mathrel{=}\ldots{}\<[E]%
\\[\blanklineskip]%
\>[B]{}\mathbf{data}\;{}\<[7]%
\>[7]{}\Conid{Exp}_{\!X}^\circ\;{}\<[14]%
\>[14]{}\mathbf{extends}\;{}\<[23]%
\>[23]{}\Conid{Exp}_{\!X}{}\<[E]%
\\
\>[B]{}\hsindent{3}{}\<[3]%
\>[3]{}\mathrel{=}{}\<[3E]%
\>[6]{}\ldots{}\<[6E]%
\\
\>[B]{}\hsindent{3}{}\<[3]%
\>[3]{}\mid {}\<[3E]%
\>[7]{}\Conid{App}_{\!X}^\circ\;{}\<[14]%
\>[14]{}\mathbf{extends}\;{}\<[23]%
\>[23]{}\Conid{App}_{\!X}\;{}\<[29]%
\>[29]{}\mathbf{by}\;{}\<[33]%
\>[33]{}\hlt{\varnothing}{}\<[E]%
\\
\>[B]{}\hsindent{3}{}\<[3]%
\>[3]{}\mid {}\<[3E]%
\>[7]{}\Conid{Let}_{\!X}^\circ\;{}\<[14]%
\>[14]{}\mathbf{extends}\;{}\<[23]%
\>[23]{}\Conid{Let}_{\!X}\;{}\<[29]%
\>[29]{}\mathbf{by}\;{}\<[33]%
\>[33]{}\hlt{\varnothing}{}\<[E]%
\\[\blanklineskip]%
\>[B]{}\newline{}\<[E]%
\ColumnHook
\end{hscode}\resethooks
\end{minipage}
\begin{minipage}[t]{0.5\textwidth}
\begin{hscode}\SaveRestoreHook
\column{B}{@{}>{\hspre}l<{\hspost}@{}}%
\column{3}{@{}>{\hspre}c<{\hspost}@{}}%
\column{3E}{@{}l@{}}%
\column{6}{@{}>{\hspre}c<{\hspost}@{}}%
\column{6E}{@{}l@{}}%
\column{7}{@{}>{\hspre}l<{\hspost}@{}}%
\column{22}{@{}>{\hspre}l<{\hspost}@{}}%
\column{E}{@{}>{\hspre}l<{\hspost}@{}}%
\>[B]{}\mathbf{data}\;{}\<[7]%
\>[7]{}\Conid{Typ}_{\!X}^\circ\;\mathbf{extends}\;\Conid{Typ}_{\!X}{}\<[E]%
\\
\>[B]{}\hsindent{3}{}\<[3]%
\>[3]{}\mathrel{=}{}\<[3E]%
\>[6]{}\ldots{}\<[6E]%
\\[\blanklineskip]%
\>[B]{}\mathbf{data}\;{}\<[7]%
\>[7]{}\Conid{Dec}_{\!X}^\circ\;\mathbf{extends}\;{}\<[22]%
\>[22]{}\Conid{Dec}_{\!X}{}\<[E]%
\\
\>[B]{}\hsindent{3}{}\<[3]%
\>[3]{}\mathrel{=}{}\<[3E]%
\>[6]{}\ldots{}\<[6E]%
\ColumnHook
\end{hscode}\resethooks
\end{minipage}

Once again, we can define the printer function by providing a printer for the
pieces of extension, and it is the same as the one for \ensuremath{\Conid{Typ}_{\!X}^\bullet}, \ensuremath{\Conid{Exp}_{\!X}^\bullet}, and
\ensuremath{\Conid{Dec}_{\!X}^\bullet}.

To recap, so far (in this subsection) we have defined
\begin{enumerate}[leftmargin=+.5in]
\setlength\itemsep{0em}
\item the extensible trees using \ensuremath{\Conid{Typ}_{\!X}}, \ensuremath{\Conid{Exp}_{\!X}}, and \ensuremath{\Conid{Dec}_{\!X}};
\item the non-decorated trees as an extension to the extensible trees using the extending data types \ensuremath{\Conid{Typ}_{\!X}^\circ}, \ensuremath{\Conid{Exp}_{\!X}^\circ}, and \ensuremath{\Conid{Dec}_{\!X}^\circ};
\item the decorated trees as an extension to the extensible trees using the extending data types \ensuremath{\Conid{Typ}_{\!X}^\bullet}, \ensuremath{\Conid{Exp}_{\!X}^\bullet} and \ensuremath{\Conid{Dec}_{\!X}^\bullet};
\item the printer for the extensible trees using \ensuremath{\Varid{printT}_{\!\!X}}, \ensuremath{\Varid{printE}_{\!\!X}}, and \ensuremath{\Varid{printD}_{\!\!X}};
\item the printer for the non-decorated trees by only defining printers for the extensions using \ensuremath{\Varid{printT}_{\!\!X}^\circ}, \ensuremath{\Varid{printE}_{\!\!X}^\circ}, and \ensuremath{\Varid{printD}_{\!\!X}^\circ};
\item the printer for the decorated trees by only defining printers for the extensions using \ensuremath{\Varid{printT}_{\!\!X}^\bullet}, \ensuremath{\Varid{printE}_{\!\!X}^\bullet}, and \ensuremath{\Varid{printD}_{\!\!X}^\bullet};
\item the type checker for the decorated trees using \ensuremath{\Varid{chkE}_{\!\!X}} and \ensuremath{\Varid{chkD}_{\!\!X}}.
\end{enumerate}

In this extensible setting, the types of the set of functions in the parser and
in the type inference engine are

\begin{hscode}\SaveRestoreHook
\column{B}{@{}>{\hspre}l<{\hspost}@{}}%
\column{E}{@{}>{\hspre}l<{\hspost}@{}}%
\>[B]{}\Varid{parseT}_{\!\!X}\mathbin{::}\Conid{String}\to \Conid{Maybe}\;\Conid{Typ}_{\!X}^\circ{}\<[E]%
\\
\>[B]{}\Varid{parseE}_{\!\!X}\mathbin{::}\Conid{String}\to \Conid{Maybe}\;\Conid{Exp}_{\!X}^\circ{}\<[E]%
\\
\>[B]{}\Varid{parseD}_{\!\!X}\mathbin{::}\Conid{String}\to \Conid{Maybe}\;\Conid{Dec}_{\!X}^\circ{}\<[E]%
\\[\blanklineskip]%
\>[B]{}\Varid{inferE}_{\!\!X}\mathbin{::}\Conid{Exp}_{\!X}^\circ\to \Conid{Exp}_{\!X}^\bullet{}\<[E]%
\\
\>[B]{}\Varid{inferD}_{\!\!X}\mathbin{::}\Conid{Dec}_{\!X}^\circ\to \Conid{Dec}_{\!X}^\bullet{}\<[E]%
\ColumnHook
\end{hscode}\resethooks

Extensibility has bought us reusability of data type declarations (and
reusability of functions in the case of printers), and also it has brought us
modularity of the definitions, in that the definitions of the decorations are
seperated from the definition of the trees.
Is this all we can do? In the next section, we identify the set of syntacically
possible forms of extensions to a data type declaration.


\section{All You Can Do}
\label{SecAllYouCanDo}
The notion of extension for a set, or for a list, is obvious: you can extend a
set by adding one or more elements to the set.  What if you are given a pair of
two sets, the set A and the set B?  There you have two forms of extensions:
extension to set A, and/or, extension to set B.  Similarly for a set of sets,
you can extend the mother set, and/or, extend the existing child sets.  However,
if you are given a pair of an atomic value (non-extensible) and a set, there
will be only one form of extensions: extending the only set. In this section, we
study notion of extension to algebraic data type declarations and generalised
algebraic data type declarations.

\subsection{Extensions in Algebraic Data Types}
Syntactically speaking, an algebraic data type declaration (ADT) in Haskell can
be seen as the structure
\begin{hscode}\SaveRestoreHook
\column{B}{@{}>{\hspre}l<{\hspost}@{}}%
\column{E}{@{}>{\hspre}l<{\hspost}@{}}%
\>[B]{}(\Conid{TyConId},\{\mskip1.5mu \Conid{VarId}\mskip1.5mu\},\{\mskip1.5mu (\Conid{ConId},[\mskip1.5mu \Conid{Type}\mskip1.5mu])\mskip1.5mu\}){}\<[E]%
\ColumnHook
\end{hscode}\resethooks
where \ensuremath{\Conid{TyConId}} represents the type constructor identifiers in Haskell, \ensuremath{\Conid{VarID}}
represents the type variable identifiers, \ensuremath{\Conid{ConId}} represents the data
constructor identifiers, \ensuremath{\Conid{Type}} represents the syntax of Haskell types, \ensuremath{\{\mskip1.5mu \anonymous \mskip1.5mu\}}
denotes sets, and \ensuremath{[\mskip1.5mu \anonymous \mskip1.5mu]} denotes lists.
The syntax of ADT
declarations, excluding the atomic parts (we noticeably consider types, except
the ones defined mutually-recursively to be atomic at this stage), consists of a
set (type variables) and a set of lists (constructors). Following a similar
reasoning as before, syntactically there are three possible forms of extensions
to an ADT declaration:\\
\indent(a) extensions to the list of fields of each data constructors,\\
\indent(b) extensions to the set of data constructors, and\\
\indent(c) extensions to the set of type parameters.

This matches exactly our specification of possible extensions to a data type in the previous section;
using the notations introduced in the previous section, we can define
extensions to an extensible algebraic data type declaration by specifying\\
\indent(a) new fields to the existing data constructors of the extensible data type,\\
\indent(b) new data constructors to the extensible data type, and\\
\indent(c) new type parameters (with alpha renaming of the existing ones, if needed).

Therefore, our notation for extensible data types is complete with respect to
the set of syntactically possible extensions to an algebraic data type
declaration.

We can go a few steps further, and consider all syntactically possible forms of
extensions to generalised algebraic data type (GADTs) declarations.

\subsection{Extensions in Generalised Algebraic Data Types}
The syntax of a generalised algebraic data type declaration (GADT) in Haskell
(ignoring the kind annotations) can be seen as the structure
\begin{hscode}\SaveRestoreHook
\column{B}{@{}>{\hspre}l<{\hspost}@{}}%
\column{E}{@{}>{\hspre}l<{\hspost}@{}}%
\>[B]{}(\Conid{TyConId},\{\mskip1.5mu \Conid{VarId}\mskip1.5mu\},\{\mskip1.5mu (\Conid{ConId},\{\mskip1.5mu \Conid{VarId}\mskip1.5mu\},\{\mskip1.5mu \Conid{Constraint}\mskip1.5mu\},[\mskip1.5mu \Conid{Type}\mskip1.5mu])\mskip1.5mu\}){}\<[E]%
\ColumnHook
\end{hscode}\resethooks
where \ensuremath{\Conid{TyConId}}, \ensuremath{\Conid{VarId}}, \ensuremath{\Conid{ConId}}, and \ensuremath{\Conid{Type}} are as before, and \ensuremath{\Conid{Constraint}}
represents the syntax of type constraints in Haskell (e.g., type class
constraints, or type equality constraints).
Following a similar reasoning as before, syntactically there are two additional
(compared to ADTs) possible forms of extensions to a GADT declaration:\\
\indent(d) extensions to the set of local type variables, and\\
\indent(e) extensions to the set of local type constraints.

Although we have not provided a syntax to describe extensions to a GADT
declaration, in theory it is straightforward to do so.

\section{GHC Can Do}
\label{SecGHCCanDo}
In this section, we present encodings of extensible data types in GHC
Haskell. Our encodings allow for all extension forms identified in the previous
section for ADTs. 

\subsection{Extensible Algebraic Data Types}
The idea behind our encoding is simple: to make an ADT declaration extensible,
introduce additional parameters to stand for possible extensions, and
instantiate these parameters per extending data types.

The simplest, yet practically na{\"i}ve, implementation of such encoding for our
running example \ensuremath{\Conid{Typ}_{\!X}} would be as follows:

\noindent
\begin{minipage}[t]{0.35\textwidth}
\begin{hscode}\SaveRestoreHook
\column{B}{@{}>{\hspre}l<{\hspost}@{}}%
\column{3}{@{}>{\hspre}c<{\hspost}@{}}%
\column{3E}{@{}l@{}}%
\column{6}{@{}>{\hspre}l<{\hspost}@{}}%
\column{E}{@{}>{\hspre}l<{\hspost}@{}}%
\>[B]{}\mathbf{extensible}\;\mathbf{data}\;\Conid{Typ}_{\!X}{}\<[E]%
\\
\>[B]{}\hsindent{3}{}\<[3]%
\>[3]{}\mathrel{=}{}\<[3E]%
\>[6]{}\Conid{Int}_{\!X}{}\<[E]%
\\
\>[B]{}\hsindent{3}{}\<[3]%
\>[3]{}\mid {}\<[3E]%
\>[6]{}\Conid{Typ}_{\!X}\overset{x}{:\!\rightarrow}\Conid{Typ}_{\!X}{}\<[E]%
\ColumnHook
\end{hscode}\resethooks
\end{minipage}
\begin{minipage}[t]{0.07\textwidth}
$$\longmapsto$$
\end{minipage}
\begin{minipage}[t]{0.4\textwidth}
\begin{hscode}\SaveRestoreHook
\column{B}{@{}>{\hspre}l<{\hspost}@{}}%
\column{3}{@{}>{\hspre}c<{\hspost}@{}}%
\column{3E}{@{}l@{}}%
\column{6}{@{}>{\hspre}l<{\hspost}@{}}%
\column{15}{@{}>{\hspre}l<{\hspost}@{}}%
\column{E}{@{}>{\hspre}l<{\hspost}@{}}%
\>[B]{}\mathbf{data}\;\Conid{Typ}_{\!X}\;\Varid{xInt}_X\;\Varid{xArr}_X\;\Varid{xTyp}_X{}\<[E]%
\\
\>[B]{}\hsindent{3}{}\<[3]%
\>[3]{}\mathrel{=}{}\<[3E]%
\>[6]{}\Conid{Int}_{\!X}\;{}\<[15]%
\>[15]{}\Varid{xInt}_X{}\<[E]%
\\
\>[B]{}\hsindent{3}{}\<[3]%
\>[3]{}\mid {}\<[3E]%
\>[6]{}(\overset{x}{:\!\rightarrow})\;{}\<[15]%
\>[15]{}\Varid{xArr}_X\;{}\<[E]%
\\
\>[15]{}(\Conid{Typ}_{\!X}\;\Varid{xInt}_X\;\Varid{xArr}_X\;\Varid{xTyp}_X)\;{}\<[E]%
\\
\>[15]{}(\Conid{Typ}_{\!X}\;\Varid{xInt}_X\;\Varid{xArr}_X\;\Varid{xTyp}_X){}\<[E]%
\\
\>[B]{}\hsindent{3}{}\<[3]%
\>[3]{}\mid {}\<[3E]%
\>[6]{}\Conid{Typ}_{\!X}\;{}\<[15]%
\>[15]{}\Varid{xTyp}_X{}\<[E]%
\\
\>[B]{}\newline{}\<[E]%
\ColumnHook
\end{hscode}\resethooks
\end{minipage}

where \ensuremath{\Varid{xInt}_X} stands for new field extensions to the constructor \ensuremath{\Conid{Int}_{\!X}}, \ensuremath{\Varid{xArr}_X}
stands for new field extensions to the constructor \ensuremath{\overset{x}{:\!\rightarrow}}, and \ensuremath{\Varid{xTyp}_X} to new
constructor extensions to \ensuremath{\Conid{Typ}_{\!X}}.

We can practically improve above encoding by adding only one (higher-order)
parameter, and project the extension parameters by a set of unique labels:

\noindent
\begin{minipage}[t]{0.35\textwidth}
\begin{hscode}\SaveRestoreHook
\column{B}{@{}>{\hspre}l<{\hspost}@{}}%
\column{3}{@{}>{\hspre}c<{\hspost}@{}}%
\column{3E}{@{}l@{}}%
\column{6}{@{}>{\hspre}l<{\hspost}@{}}%
\column{E}{@{}>{\hspre}l<{\hspost}@{}}%
\>[B]{}\mathbf{extensible}\;\mathbf{data}\;\Conid{Typ}_{\!X}{}\<[E]%
\\
\>[B]{}\hsindent{3}{}\<[3]%
\>[3]{}\mathrel{=}{}\<[3E]%
\>[6]{}\Conid{Int}_{\!X}{}\<[E]%
\\
\>[B]{}\hsindent{3}{}\<[3]%
\>[3]{}\mid {}\<[3E]%
\>[6]{}\Conid{Typ}_{\!X}\overset{x}{:\!\rightarrow}\Conid{Typ}_{\!X}{}\<[E]%
\ColumnHook
\end{hscode}\resethooks
\end{minipage}
\begin{minipage}[t]{0.07\textwidth}
$$\longmapsto$$
\end{minipage}
\begin{minipage}[t]{0.4\textwidth}
\begin{hscode}\SaveRestoreHook
\column{B}{@{}>{\hspre}l<{\hspost}@{}}%
\column{3}{@{}>{\hspre}c<{\hspost}@{}}%
\column{3E}{@{}l@{}}%
\column{6}{@{}>{\hspre}l<{\hspost}@{}}%
\column{15}{@{}>{\hspre}l<{\hspost}@{}}%
\column{28}{@{}>{\hspre}l<{\hspost}@{}}%
\column{E}{@{}>{\hspre}l<{\hspost}@{}}%
\>[B]{}\mathbf{data}\;\Conid{Typ}_{\!X}\;\xi{}\<[E]%
\\
\>[B]{}\hsindent{3}{}\<[3]%
\>[3]{}\mathrel{=}{}\<[3E]%
\>[6]{}\Conid{Int}_{\!X}\;{}\<[15]%
\>[15]{}(\xi\;\text{\tt \char34 IntX\char34}){}\<[E]%
\\
\>[B]{}\hsindent{3}{}\<[3]%
\>[3]{}\mid {}\<[3E]%
\>[6]{}(\overset{x}{:\!\rightarrow})\;{}\<[15]%
\>[15]{}(\xi\;\text{\tt \char34 ArrX\char34})\;{}\<[28]%
\>[28]{}(\Conid{Typ}_{\!X}\;\xi)\;(\Conid{Typ}_{\!X}\;\xi){}\<[E]%
\\
\>[B]{}\hsindent{3}{}\<[3]%
\>[3]{}\mid {}\<[3E]%
\>[6]{}\Conid{Typ}_{\!X}\;{}\<[15]%
\>[15]{}(\xi\;\text{\tt \char34 TypX\char34}){}\<[E]%
\\
\>[B]{}\newline{}\<[E]%
\ColumnHook
\end{hscode}\resethooks
\end{minipage}

The data types extending \ensuremath{\Conid{Typ}_{\!X}} can be defined by setting the
parameters. However, in the latter, more compact, variant we need to instantiate
a higher-order (indexed) parameter in Haskell, and we can choose to do so either
by GADTs, or by data families. For example, the extending data type \ensuremath{\Conid{Typ}_{\!X}^\bullet} from
earlier can be encoded as follows:

\noindent
\begin{minipage}[t]{0.4\textwidth}
\begin{hscode}\SaveRestoreHook
\column{B}{@{}>{\hspre}l<{\hspost}@{}}%
\column{3}{@{}>{\hspre}c<{\hspost}@{}}%
\column{3E}{@{}l@{}}%
\column{7}{@{}>{\hspre}l<{\hspost}@{}}%
\column{16}{@{}>{\hspre}l<{\hspost}@{}}%
\column{25}{@{}>{\hspre}l<{\hspost}@{}}%
\column{33}{@{}>{\hspre}l<{\hspost}@{}}%
\column{37}{@{}>{\hspre}l<{\hspost}@{}}%
\column{E}{@{}>{\hspre}l<{\hspost}@{}}%
\>[B]{}\mathbf{data}\;{}\<[7]%
\>[7]{}\Conid{Typ}_{\!X}^\bullet\;\mathbf{extends}\;\Conid{Typ}_{\!X}{}\<[E]%
\\
\>[B]{}\hsindent{3}{}\<[3]%
\>[3]{}\mathrel{=}{}\<[3E]%
\>[7]{}\Conid{Typ}_{\!X}^\bullet:\!\!*\!\!:\!^\bullet\Conid{Typ}_{\!X}^\bullet{}\<[E]%
\\
\>[B]{}\hsindent{3}{}\<[3]%
\>[3]{}\mid {}\<[3E]%
\>[7]{}\Conid{Int}_{\!X}^\bullet\;{}\<[16]%
\>[16]{}\mathbf{extends}\;{}\<[25]%
\>[25]{}\Conid{Int}_{\!X}\;{}\<[33]%
\>[33]{}\mathbf{by}\;{}\<[37]%
\>[37]{}\varnothing{}\<[E]%
\\
\>[B]{}\hsindent{3}{}\<[3]%
\>[3]{}\mid {}\<[3E]%
\>[7]{}(\overset{x}{:\!\rightarrow}^\bullet)\;{}\<[16]%
\>[16]{}\mathbf{extends}\;{}\<[25]%
\>[25]{}(\overset{x}{:\!\rightarrow})\;{}\<[33]%
\>[33]{}\mathbf{by}\;{}\<[37]%
\>[37]{}\varnothing{}\<[E]%
\ColumnHook
\end{hscode}\resethooks
\end{minipage}
\begin{minipage}[t]{0.05\textwidth}
$\ $
\newline
$\ $
\newline
$\ $
\newline
$\ $
\newline
$\ $
\newline
$\longmapsto\\$
\end{minipage}
\begin{minipage}[t]{0.5\textwidth}
\begin{hscode}\SaveRestoreHook
\column{B}{@{}>{\hspre}l<{\hspost}@{}}%
\column{3}{@{}>{\hspre}l<{\hspost}@{}}%
\column{20}{@{}>{\hspre}c<{\hspost}@{}}%
\column{20E}{@{}l@{}}%
\column{23}{@{}>{\hspre}l<{\hspost}@{}}%
\column{E}{@{}>{\hspre}l<{\hspost}@{}}%
\>[B]{}\mathbf{type}\;\Conid{Typ}_{\!X}^\bullet\mathrel{=}\Conid{Typ}_{\!X}\;\Conid{Ext}{}\<[E]%
\\[\blanklineskip]%
\>[B]{}\mathbf{data}\;\mathbf{family}\;\Conid{Ext}\;(\Varid{label}\mathbin{::}\Conid{Symbol})\mathbin{::}\mathbin{*}{}\<[E]%
\\[\blanklineskip]%
\>[B]{}\mathbf{data}\;\mathbf{instance}\;\Conid{Ext}\;\text{\tt \char34 TypX\char34}{}\<[E]%
\\
\>[B]{}\hsindent{3}{}\<[3]%
\>[3]{}\mathrel{=}\Conid{Typ}_{\!X}^\bullet:\!\!*\!\!:\!^\prime\Conid{Typ}_{\!X}^\bullet{}\<[E]%
\\
\>[B]{}\mathbf{data}\;\mathbf{instance}\;\Conid{Ext}\;\text{\tt \char34 IntX\char34}\mathrel{=}\Conid{NoneI}{}\<[E]%
\\
\>[B]{}\mathbf{data}\;\mathbf{instance}\;\Conid{Ext}\;\text{\tt \char34 ArrX\char34}\mathrel{=}\Conid{NoneA}{}\<[E]%
\\[\blanklineskip]%
\>[B]{}\mathbf{pattern}\;\Varid{m}:\!\!*\!\!:\!^\bullet\Varid{n}{}\<[20]%
\>[20]{}\mathrel{=}{}\<[20E]%
\>[23]{}\Conid{Typ}_{\!X}\;(\Varid{m}:\!\!*\!\!:\!^\prime\Varid{n}){}\<[E]%
\\
\>[B]{}\mathbf{pattern}\;\Conid{Int}_{\!X}^\bullet{}\<[20]%
\>[20]{}\mathrel{=}{}\<[20E]%
\>[23]{}\Conid{Int}_{\!X}\;\Conid{NoneI}{}\<[E]%
\\
\>[B]{}\mathbf{pattern}\;\Varid{m}\overset{x}{:\!\rightarrow}^\bullet\Varid{n}{}\<[20]%
\>[20]{}\mathrel{=}{}\<[20E]%
\>[23]{}(\overset{x}{:\!\rightarrow})\;\Conid{NoneA}\;\Varid{m}\;\Varid{n}{}\<[E]%
\\
\>[B]{}\newline{}\<[E]%
\ColumnHook
\end{hscode}\resethooks
\end{minipage}

The process of translating from our syntax to the underlying encoding is as follows.

\noindent
\begin{minipage}[t]{0.4\textwidth}
Declarations:
\begin{hscode}\SaveRestoreHook
\column{B}{@{}>{\hspre}l<{\hspost}@{}}%
\column{3}{@{}>{\hspre}c<{\hspost}@{}}%
\column{3E}{@{}l@{}}%
\column{7}{@{}>{\hspre}l<{\hspost}@{}}%
\column{E}{@{}>{\hspre}l<{\hspost}@{}}%
\>[B]{}\mathbf{extensible}\;\mathbf{data}\;\Conid{T'}\;\alpha_1\ldots\alpha_n{}\<[E]%
\\
\>[B]{}\hsindent{3}{}\<[3]%
\>[3]{}\mathrel{=}{}\<[3E]%
\>[7]{}\ldots{}\<[E]%
\\
\>[B]{}\hsindent{3}{}\<[3]%
\>[3]{}\mid {}\<[3E]%
\>[7]{}\Conid{C}_{\Varid{i}}\ldots\Conid{T}_{i,j}\ldots{}\<[E]%
\\
\>[B]{}\hsindent{3}{}\<[3]%
\>[3]{}\mid {}\<[3E]%
\>[7]{}\ldots{}\<[E]%
\ColumnHook
\end{hscode}\resethooks
\end{minipage}
\begin{minipage}[t]{0.05\textwidth}
$\ $
\newline
$\ $
\newline
$\ $
\newline
$\longmapsto\\$
\end{minipage}
\begin{minipage}[t]{0.5\textwidth}
\begin{hscode}\SaveRestoreHook
\column{B}{@{}>{\hspre}l<{\hspost}@{}}%
\column{3}{@{}>{\hspre}c<{\hspost}@{}}%
\column{3E}{@{}l@{}}%
\column{6}{@{}>{\hspre}l<{\hspost}@{}}%
\column{11}{@{}>{\hspre}l<{\hspost}@{}}%
\column{16}{@{}>{\hspre}l<{\hspost}@{}}%
\column{70}{@{}>{\hspre}c<{\hspost}@{}}%
\column{70E}{@{}l@{}}%
\column{E}{@{}>{\hspre}l<{\hspost}@{}}%
\>[B]{}\mathbf{data}\;\Conid{T'}\;\xi\;\alpha_1\ldots\alpha_n{}\<[E]%
\\
\>[B]{}\hsindent{3}{}\<[3]%
\>[3]{}\mathrel{=}{}\<[3E]%
\>[6]{}\Conid{T'}\;{}\<[11]%
\>[11]{}(\xi\;{}\<[16]%
\>[16]{}\text{\tt \char34 T'\char34}){}\<[E]%
\\
\>[B]{}\hsindent{3}{}\<[3]%
\>[3]{}\mid {}\<[3E]%
\>[6]{}\ldots{}\<[E]%
\\
\>[B]{}\hsindent{3}{}\<[3]%
\>[3]{}\mid {}\<[3E]%
\>[6]{}\Conid{C}_{\Varid{i}}\;{}\<[11]%
\>[11]{}(\xi\;{}\<[16]%
\>[16]{}\text{\tt \char34 C$_i$\char34})\ldots\llbracket\Conid{T}_{i,j}\rrbracket_\xi{}\<[70]%
\>[70]{}\ldots{}\<[70E]%
\\
\>[B]{}\hsindent{3}{}\<[3]%
\>[3]{}\mid {}\<[3E]%
\>[6]{}\ldots{}\<[E]%
\\
\>[B]{}\newline{}\<[E]%
\ColumnHook
\end{hscode}\resethooks
\end{minipage}
\begin{centering}
\begin{hscode}\SaveRestoreHook
\column{B}{@{}>{\hspre}l<{\hspost}@{}}%
\column{12}{@{}>{\hspre}l<{\hspost}@{}}%
\column{15}{@{}>{\hspre}l<{\hspost}@{}}%
\column{28}{@{}>{\hspre}c<{\hspost}@{}}%
\column{28E}{@{}l@{}}%
\column{31}{@{}>{\hspre}l<{\hspost}@{}}%
\column{37}{@{}>{\hspre}l<{\hspost}@{}}%
\column{E}{@{}>{\hspre}l<{\hspost}@{}}%
\>[B]{}\llbracket\;{}\<[12]%
\>[12]{}\Conid{T}\;{}\<[15]%
\>[15]{}\rrbracket_\xi{}\<[28]%
\>[28]{}\mathrel{=}{}\<[28E]%
\>[31]{}\Conid{T}\;\xi{}\<[37]%
\>[37]{}\ \ \text{   if $\Conid{T}$ is extensible}{}\<[E]%
\\
\>[B]{}\llbracket\;{}\<[12]%
\>[12]{}\Conid{T}\;{}\<[15]%
\>[15]{}\rrbracket_\xi{}\<[28]%
\>[28]{}\mathrel{=}{}\<[28E]%
\>[31]{}\Conid{T}{}\<[37]%
\>[37]{}\ \ \text{   if $\Conid{T}$ is not extensible}{}\<[E]%
\ColumnHook
\end{hscode}\resethooks
\end{centering}

\noindent
\begin{minipage}[t]{0.4\textwidth}
Types:
\begin{hscode}\SaveRestoreHook
\column{B}{@{}>{\hspre}l<{\hspost}@{}}%
\column{E}{@{}>{\hspre}l<{\hspost}@{}}%
\>[B]{}\Conid{T}_{\Conid{K}}\;\Conid{T}_{\mathrm{1}}\ldots\Conid{T}_{\Varid{n}}\;\!\!\boldsymbol{\oplus}\!\!\;\Conid{T}_\xi{}\<[E]%
\ColumnHook
\end{hscode}\resethooks
Patterns:
\begin{hscode}\SaveRestoreHook
\column{B}{@{}>{\hspre}l<{\hspost}@{}}%
\column{E}{@{}>{\hspre}l<{\hspost}@{}}%
\>[B]{}\Conid{C}\;\Conid{P}_{\mathrm{1}}\ldots\Conid{P}_{\Varid{n}}\;\!\!\boldsymbol{\oplus}\!\!\;\Conid{P'}{}\<[E]%
\ColumnHook
\end{hscode}\resethooks
Data Constructors:
\begin{hscode}\SaveRestoreHook
\column{B}{@{}>{\hspre}l<{\hspost}@{}}%
\column{E}{@{}>{\hspre}l<{\hspost}@{}}%
\>[B]{}\Conid{C}\;\Conid{M}_{\mathrm{1}}\ldots\Conid{M}_{\Varid{n}}\;\!\!\boldsymbol{\oplus}\!\!\;\Conid{M'}{}\<[E]%
\ColumnHook
\end{hscode}\resethooks
\end{minipage}
\begin{minipage}[t]{0.05\textwidth}
$\ $
\newline
$\longmapsto\\$
$\longmapsto\\$
$\longmapsto\\$
\end{minipage}
\begin{minipage}[t]{0.5\textwidth}
$ $
\begin{hscode}\SaveRestoreHook
\column{B}{@{}>{\hspre}l<{\hspost}@{}}%
\column{E}{@{}>{\hspre}l<{\hspost}@{}}%
\>[B]{}\Conid{T}_{\Conid{K}}\;\Conid{T}_\xi\;\Conid{T}_{\mathrm{1}}\ldots\Conid{T}_{\Varid{n}}{}\<[E]%
\ColumnHook
\end{hscode}\resethooks
$ $
\begin{hscode}\SaveRestoreHook
\column{B}{@{}>{\hspre}l<{\hspost}@{}}%
\column{E}{@{}>{\hspre}l<{\hspost}@{}}%
\>[B]{}\Conid{C}\;\Conid{P'}\;\Conid{P}_{\mathrm{1}}\ldots\Conid{P}_{\Varid{n}}{}\<[E]%
\ColumnHook
\end{hscode}\resethooks
$ $
\begin{hscode}\SaveRestoreHook
\column{B}{@{}>{\hspre}l<{\hspost}@{}}%
\column{E}{@{}>{\hspre}l<{\hspost}@{}}%
\>[B]{}\Conid{C}\;\Conid{M'}\;\Conid{M}_{\mathrm{1}}\ldots\Conid{M}_{\Varid{n}}{}\<[E]%
\ColumnHook
\end{hscode}\resethooks
\end{minipage}

\noindent
\begin{minipage}[t]{0.4\textwidth}
Extensions:
\begin{hscode}\SaveRestoreHook
\column{B}{@{}>{\hspre}l<{\hspost}@{}}%
\column{3}{@{}>{\hspre}c<{\hspost}@{}}%
\column{3E}{@{}l@{}}%
\column{6}{@{}>{\hspre}l<{\hspost}@{}}%
\column{7}{@{}>{\hspre}l<{\hspost}@{}}%
\column{11}{@{}>{\hspre}c<{\hspost}@{}}%
\column{11E}{@{}l@{}}%
\column{12}{@{}>{\hspre}l<{\hspost}@{}}%
\column{13}{@{}>{\hspre}l<{\hspost}@{}}%
\column{16}{@{}>{\hspre}l<{\hspost}@{}}%
\column{25}{@{}>{\hspre}l<{\hspost}@{}}%
\column{E}{@{}>{\hspre}l<{\hspost}@{}}%
\>[B]{}\mathbf{data}\;{}\<[7]%
\>[7]{}\Conid{T'}{}\<[11]%
\>[11]{}\ldots{}\<[11E]%
\>[16]{}\alpha_i{}\<[25]%
\>[25]{}\ldots\mathbf{extends}{}\<[E]%
\\
\>[7]{}(\Conid{T}{}\<[11]%
\>[11]{}\ldots{}\<[11E]%
\>[16]{}\beta_j{}\<[25]%
\>[25]{}\ldots){}\<[E]%
\\
\>[B]{}\hsindent{3}{}\<[3]%
\>[3]{}\mathrel{=}{}\<[3E]%
\>[6]{}\ldots{}\<[E]%
\\
\>[B]{}\hsindent{3}{}\<[3]%
\>[3]{}\mid {}\<[3E]%
\>[6]{}\Conid{C}^\prime_{\Varid{i}^\prime}\;{}\<[13]%
\>[13]{}\mathbf{extends}\;\Conid{C}_{\Varid{i}^\prime}\;\mathbf{by}\;{}\<[E]%
\\
\>[6]{}\hsindent{6}{}\<[12]%
\>[12]{}\Conid{T}_{i^\prime,1}\ldots\Conid{T}_{i^\prime,m}{}\<[E]%
\\
\>[B]{}\hsindent{3}{}\<[3]%
\>[3]{}\mid {}\<[3E]%
\>[6]{}\ldots{}\<[E]%
\\
\>[B]{}\hsindent{3}{}\<[3]%
\>[3]{}\mid {}\<[3E]%
\>[6]{}\Conid{C}^\prime_{\Varid{j}^\prime}\;\Conid{T}_{j^\prime,1}\ldots\Conid{T}_{j^\prime,n}{}\<[E]%
\\
\>[B]{}\hsindent{3}{}\<[3]%
\>[3]{}\mid {}\<[3E]%
\>[6]{}\ldots{}\<[E]%
\ColumnHook
\end{hscode}\resethooks
\end{minipage}
\begin{minipage}[t]{0.05\textwidth}
$\ $
\newline
$\ $
\newline
$\ $
\newline
$\longmapsto\\$
\end{minipage}
\begin{minipage}[t]{0.5\textwidth}
\begin{hscode}\SaveRestoreHook
\column{B}{@{}>{\hspre}l<{\hspost}@{}}%
\column{3}{@{}>{\hspre}l<{\hspost}@{}}%
\column{6}{@{}>{\hspre}l<{\hspost}@{}}%
\column{13}{@{}>{\hspre}l<{\hspost}@{}}%
\column{16}{@{}>{\hspre}l<{\hspost}@{}}%
\column{22}{@{}>{\hspre}l<{\hspost}@{}}%
\column{E}{@{}>{\hspre}l<{\hspost}@{}}%
\>[B]{}\mathbf{type}\;\Conid{T'}\ldots\alpha_i\ldots{}\<[E]%
\\
\>[B]{}\mathrel{=}\Conid{T}\;\Conid{Ext}^u\ldots\beta_j\ldots{}\<[E]%
\\[\blanklineskip]%
\>[B]{}\mathbf{data}\;\mathbf{family}\;\Conid{Ext}^u\;(\Varid{label}\mathbin{::}\Conid{Symbol})\mathbin{::}\mathbin{*}{}\<[E]%
\\[\blanklineskip]%
\>[B]{}\mathbf{data}\;\mathbf{instance}\;{}\<[16]%
\>[16]{}\Conid{Ext}^u\;\text{\tt \char34 T\char34}{}\<[E]%
\\
\>[B]{}\hsindent{3}{}\<[3]%
\>[3]{}\mathrel{=}{}\<[6]%
\>[6]{}\ldots{}\<[E]%
\\
\>[B]{}\hsindent{3}{}\<[3]%
\>[3]{}\mid {}\<[6]%
\>[6]{}\Conid{C}^\prime_{\Varid{j}^\prime}\;\Conid{T}_{j^\prime,1}\ldots\Conid{T}_{j^\prime,n}{}\<[E]%
\\
\>[B]{}\hsindent{3}{}\<[3]%
\>[3]{}\mid {}\<[6]%
\>[6]{}\ldots{}\<[E]%
\\[\blanklineskip]%
\>[B]{}\ldots{}\<[E]%
\\
\>[B]{}\mathbf{data}\;\mathbf{instance}\;{}\<[16]%
\>[16]{}\Conid{Ext}^u\;{}\<[22]%
\>[22]{}\text{\tt \char34 C$^\prime_{i^\prime}$\char34}{}\<[E]%
\\
\>[B]{}\hsindent{3}{}\<[3]%
\>[3]{}\mathrel{=}\Conid{C}^{\prime u}_{\Varid{i}^\prime}{}\<[13]%
\>[13]{}\ldots\Conid{T}_{i^\prime,k}\ldots{}\<[E]%
\\
\>[B]{}\mathbf{pattern}\;\Conid{C}^\prime_{\Varid{i}^\prime}\;\Varid{x}_{\mathrm{1}}\ldots\Varid{x}_{\Varid{m}}\;\Varid{y}_{\mathrm{1}}\ldots\Varid{y}_{\Varid{k}}{}\<[E]%
\\
\>[B]{}\hsindent{3}{}\<[3]%
\>[3]{}\mathrel{=}\Conid{C}_{\Varid{i}^\prime}\;(\Conid{C}^{\prime u}_{\Varid{i}^\prime}\;\Varid{x}_{\mathrm{1}}\ldots\Varid{x}_{\Varid{m}})\;\Varid{y}_{\mathrm{1}}\ldots\Varid{y}_{\Varid{k}}{}\<[E]%
\\
\>[B]{}\ldots{}\<[E]%
\\
\>[B]{}\newline{}\<[E]%
\ColumnHook
\end{hscode}\resethooks
\end{minipage}

We write \ensuremath{\Conid{Id}^u}, or \ensuremath{\Varid{id}^u}, to represent unique generated names. This encoding
naturally scales to mutually-recursive definitions by using the same extension
data family.

\subsection{Extensible Generalised Algebraic Data Types}
So far, we have considered encoding of extensible algebraic data type
declarations. We can extend our results to generalised algebraic data types.

\bibliographystyle{abbrvnamed}
\bibliography{Paper}

\end{document}